\documentclass[12pt,a4paper]{article}
\pdfoutput=1
\usepackage{jheppub}
\usepackage[T1]{fontenc} 
\usepackage{epsfig} 
\usepackage{amscd}
\usepackage{verbatim}
\usepackage{latexsym}
\usepackage{amsfonts,amsthm,amsmath,mathtools}
\usepackage{color}
\usepackage{amssymb,stmaryrd,textcomp,enumerate,bbm}
\usepackage{graphicx}
\usepackage{verbatim}
\usepackage[utf8]{inputenc}

\usepackage[utf8]{inputenc}
\newcommand{\mathsym}[1]{{}}

\newcommand{\be}{\begin{equation}}
\newcommand{\ee}{\end{equation}}

\newcommand{\bea}{\begin{eqnarray}}
\newcommand{\eea}{\end{eqnarray}}

\title{{Betti multiplets, flows across dimensions and $c$-extremization
}}
\author[a,b]{Antonio Amariti}
\author[c]{, Chiara Toldo}
\affiliation[a]{Physics Department, The City College of the CUNY,
160 Convent Avenue, New York, NY 10031, USA}
\affiliation[b]{Albert Einstein Center for Fundamental Physics, Institute for Theoretical Physics, University of Bern,
Sidlerstrasse 5, Bern, ch-3012, Switzerland}
\affiliation[c]{Columbia University in the City of New York,  538 West 120th Street, 704 Pupin Hall, MC 5225, New York, NY 10027, USA}

\emailAdd{amariti@itp.unibe.ch,ct2673@columbia.edu}

\abstract{
We consider 4d $\mathcal{N}=1$ SCFTs, 
topologically twisted on compact constant curvature Riemann surfaces, giving rise to
2d $\mathcal{N}=(0,2)$ SCFTs. 
 The exact R-current of these 2d SCFT
 extremizes the central charge $c_{2d}$, similarly to the 4d picture,
 where the exact R-current maximizes the central charge $a_{4d}$.
There are global currents that do not mix with the R-current in 4d but their mixing becomes non trivial in 2d.  
In this paper we study the holographic dual of this process by analyzing a 5d $\mathcal{N}=2$  truncation of $T^{1,1}$
with one Betti vector multiplet, dual to the baryonic current on the CFT side. The 
holographic realization of the flow across dimensions connects AdS$_5$ to AdS$_3$ vacua in the supergravity picture. We verify the existence of the flow to AdS$_3$ solutions
and we retrieve the field theory results for the mixing of the Betti vector
with the graviphoton. Moreover, we extract the central charge 
from the Brown-Henneaux formula, matching with the results obtained in field theory. We develop a general formalism to obtain the central charge of a 2d SCFT
 from  5d $\mathcal{N}=2$ gauged supergravity with a generic number of vector multiplets, showing
that its extremization corresponds to an attractor mechanism for the scalars in the supergravity picture.}

\begin{document}

\maketitle

\section{Introduction}

A useful mechanism to construct SCFTs in $d$-dimensions consists in compactifying a 
$(d+n)$-dimensional superconformal theory on a compact $n$-dimensional manifold $\mathcal{M}_n$.
By turning on background magnetic fluxes for the global symmetries some amount of supersymmetry of the $d+n$ dimensional theory can be preserved in $d$ dimensions.
This procedure has been developed in \cite{Witten:1988xj} and it is usually referred as (partial) topological twist. 

The topological twists does not guarantee the existence of a superconformal symmetry in the lower dimensional theory. A powerful hint for the existence of such symmetry is provided by the AdS/CFT correspondence. Let us consider a four dimensional $\mathcal{N}=1$ SCFT, with an AdS$_5$ gravity dual description, and compactify such model on a two dimensional compact Riemann surface of genus $\mathfrak{g}$, $\Sigma_{\mathfrak{g}}$.
The existence of a $\mathcal{N}=(0,2)$ SCFT in two dimensions can be inferred from the existence of the gravitational flow from AdS$_5$ to AdS$_3 \times \Sigma_g$.
Finding an explicit gravitational flow usually requires the knowledge of the full 10 dimensional 
geometry. A  rich variety of examples has been worked out in the literature, see for example 
 \cite{Bershadsky:1995vm,Bershadsky:1995qy,Maldacena:2000mw,Johansen:2003hw,Kapustin:2006hi,Donos:2008ug,Benini:2013cda,Donos:2014eua,Baggio:2014hua,Bea:2015fja,Benini:2015bwz,Schafer-Nameki:2016cfr,Apruzzi:2016iac,Apruzzi:2016nfr}. For classifications of AdS$_3$ solutions with uplift in 10d and 11d see \cite{Kim:2005ez} and \cite{Figueras:2007cn}.

In particular in \cite{Benini:2015bwz} the authors studied the gravity dual of an infinite set of quiver gauge theories, namely $Y^{pq}$ theories \cite{Benvenuti:2004dy}. They arise by the worldvolume gauge theory living on a stack of $N$ D3 branes probing the tip of a three dimensional Calabi-Yau cone with a five dimensional Sasaki Einstein (SE$_5$) base, in this case a $Y^{pq}$ manifold \cite{Gauntlett:2004yd}.
Such theories are $\mathcal{N}=1$ quiver gauge theories in 4d and, by turning on background magnetic fluxes for the global symmetries, it has been shown that in two dimensions they can flow to theories with $\mathcal{N}=(0,2)$ supersymmetry.
The possible existence of a superconformal phase has been shown for suitable choices of the quantized fluxes. This has been checked by studying the central charge on the field theory side and 
matching with the solutions on the dual gravity  side.

This analysis revealed an interesting feature regarding the exact R-current of the
2d $\mathcal{N}=(0,2)$ theories.
As in the 4d case indeed the exact R-current is a linear combinations of the UV R-current and
of the abelian global symmetries.
In 4d the mixing coefficients can be obtained  by maximizing the conformal anomaly $a_{4d}$ \cite{Intriligator:2003jj}.
In the $Y^{pq}$ case there the are in general two possible sources of mixing: the mesonic symmetries and the baryonic ones. A feature of these models is that the baryonic symmetry does not contribute to the mixing \footnote{See \cite{Bertolini:2004xf} for a discussion of the mixing of baryonic symmetries in the $Y^{21}$ model, and for example \cite{Butti:2005vn} for a general discussion on the mixing in SE$_5$ manifolds.}, namely the exact R-current in 4d is a linear combination of the UV R-current and of the 
(mesonic) flavor symmetries \emph{only}. 
However, after the compactification of the theories on $\Sigma_g$ the baryonic symmetry in general  contributes to the exact R-current of the 2d theory  \cite{Benini:2015bwz}. The
mixing coefficient is computed by extremizing the two dimensional central charge.

This mechanism can be captured on the supergravity side, by considering a 5d model arising from a consistent truncation of 10d supergravity.
The baryonic symmetry is associated to the presence of a so called Betti vector multiplet in the spectrum of the five dimensional $\mathcal{N}=2$ gauged supergravity
\footnote{See also \cite{Schmude:2016bqp} for a recent analysis of the Betti multiplets in these models.}.
Hence by considering a truncation with a graviton and a Betti vector multiplet one should see  the absence of mixing on the 5d theory and the presence of such mixing after the flow to 3d.

In this paper we check this idea by studying an explicit truncation of the $Y^{10}=T^{1,1}$ theory, dual to the Klebanov Witten theory \cite{Klebanov:1998hh}. The supergravity truncation was provided in \cite{Cassani:2010na} (see also \cite{Bena:2010pr}) and it requires in particular the presence of hypermultiplets and massive vectors. We develop the necessary formalism in 5d $\mathcal{N}=2$ gauged supergravity to study the flow across dimensions in presence of background fluxes and obtain the 2d central charge 
in terms of the sections, the isometries and the fluxes.
Notice that the relation between $c$-extremization and the $\mathcal{N}=2$  gauged supergravity BPS flow equations has already been proposed in \cite{Karndumri:2013iqa}, where the $STU$ truncation (one graviton and two massless vector multiplets) was taken into consideration.
Here we consider a more general setup, without fixing the number of vector multiplets and discussing the analysis in presence of  hypermultiplets.

The paper is organized as follows. In section \ref{sec:top} we review the compactification of 4d $\mathcal{N}=1$ SCFTs
 on compact Riemann surfaces by a partial topological twist.
 We focus on the Klebanov Witten theory, reproducing the formula for $c_{2d}$ in presence of a twist along the baryonic symmetry.
  In section \ref{sec:sugra35} we study the flows across dimensions from the perspective
 of gauged supergravity in presence of a graviton and $n_V$ vector multiplets.
  We concentrate on the analysis of  the central charges
 of the dual SCFTs from the gravitational perspective. We derive a general formula for $c_{2d}$ in terms of the gravitational quantities and show that its extremization corresponds to an attractor mechanism in supergravity.
 We conclude the analysis by discussing how the formalism is modified by the presence of
 matter hypermultiplets in subsection \ref{hyperlagrange}. Sections \ref{sec:n4}, \ref{sec:conifoldBetti},  \ref{M5} are devoted to the discussion of specific examples this formulation can be applied to.
 In section \ref{sec:n4} we apply the formalism to the compactification of $\mathcal{N}=4$ SYM
with a generic twist, breaking supersymmetry to $\mathcal{N}=(0,2)$ in 2d, and match our
formula with the results of \cite{Benini:2013cda}.
In section \ref{sec:conifoldBetti} we discuss the main result of our paper. We study a truncation of $T^{1,1}$
in presence of a Betti vector multiplet. For consistency the truncation requires also the presence of a massive vector multiplet and hypermultiplets.
We study the compactification on the Riemann surface on the supergravity side, 
finding the holographic RG flow to the AdS$_3$ vacuum.
Then we compute $c_{2d}$ from supergravity, with the techniques
developed in section \ref{sec:sugra35},
reproducing the field theory results and showing the mixing of the Betti multiplet with the the graviphoton.
In section \ref{M5} we apply our formalism to another example, corresponding to the
truncation of 7d $\mathcal{N} = 4$ $SO(5)$ gauged supergravity reduced on a Riemann surface.
This truncation is dual to the infinite family of 4d $\mathcal{N}=1$ 
SCFTs  associated to M5 branes worked out in \cite{Bah:2012dg}.
In this case we recover the large $N$ results already obtained in 
\cite{Benini:2013cda,Baggio:2014hua}.
%In section \ref{sec:LSfp} we study the last example, the $\mathcal{N}=2$
%truncation of 5d maximal gauged supergravity
%worked out in \cite{Bobev:2014jva}, corresponding to the holographic dual description
%of the $\mathcal{N}=1^*$ Leigh-Strassler fixed point \cite{Leigh:1995ep}.
%By reformulating this theory in terms of $\mathcal{N}=2$
%gauged supergravity and applying our formalism we recover the 2d
%central charge obtained by the twisted compactification 
%of this model on an hyperbolic Riemann surface.
In section \ref{sec:Conclusions} we conclude and discuss possible lines of research.
Two appendices clarify notations and details of the calculations. In appendix \ref{AppA} we provide a review of gauged supergravity,
fixing the conventions used in the paper.
In appendix \ref{variation} we provide the details of derivation of the BPS equations for the truncation
of $T^{1,1}$ with the Betti vector multiplet on a constant curvature Riemann surface.

{\bf Note Added}: When finishing this paper we noticed \cite{Klemm:2016kxw}  on the ArXiv, where 
attractor equations for extremal black strings in presence of hypermultiplets have been derived via rewriting the action in a sum of squares, and a prescription to compute $c_{2d}$ from 5d $\mathcal{N}=2$ gauged supergravity is provided.

\section{Flowing from 4d $\mathcal{N}=1$  to 2d $\mathcal{N}=(0,2)$  SCFTs}
\label{sec:top}

In this section we discuss some aspects of the flow across dimensions 
discussed in \cite{Benini:2015bwz}. More concretely we sketchily review the 
derivation of the central charge $c_{2d}$ from
4d SCFTs on two dimensional compact Riemann surfaces
with a partial topological twist along the abelian symmetries.

Consider a four dimensional $\mathcal{N}=1$ SCFT. At the fixed point the exact R-current
is determined by a linear combination of the abelian flavor symmetries $J_{F_i}$ and of the 
UV R-current, $R_0$
\begin{equation}
R = R_0 + \alpha_i J_{F_i}\,.
\end{equation}
The exact coefficients $\alpha_i$  of such mixing are determined by maximizing the central charge $a_{4d}$ \cite{Intriligator:2003jj}
\begin{equation}
\label{a4d}
a_{4d} = \frac{3}{32} ( 3\text{Tr} R^3 - \text{Tr} R)\,.
\end{equation}
Such a superconformal theory can be compactified on a genus $\mathfrak{g}$ compact
Riemann surface
$\Sigma_\mathfrak{g}$: 
some supersymmetries can be preserved if some background magnetic fluxes
are turned on, canceling the spin connection on the curved manifold.
In general one can turn on such background fluxes for all the non anomalous 
global abelian symmetries of the 4d theory \footnote{One can also twist along 
non-abelian symmetries   along their Cartan subgroup. This
procedure breaks such symmetries to $U(1)$, allowing them to mix with the
R-current in two dimensions.}.

Define $T_i$ as the generator of the $i$-th abelian symmetry.
One turns on the external fluxes as $F_i = T_i vol_{\Sigma_\mathfrak{g}}$,
where $vol_{\Sigma_{\mathfrak{g}}}$ is the volume form, normalized such that
$\int d vol_{\Sigma_{\mathfrak{g}}} = 2 \pi \eta_{\Sigma}$ and
$\eta_{\Sigma} = 2 |\mathfrak{g}-1|$ for $\mathfrak{g}\neq 1$ and $\eta_{\Sigma}=1$ for $\mathfrak{g}=1$. 
Such fluxes are subject to opportune Dirac
quantization conditions (see \cite{Benini:2013cda,Benini:2015bwz} for details).

So far the discussion has been general, and it can be applied to any 4d SCFT.
In \cite{Benini:2015bwz} an infinite family of 4d SCFTs has been considered. This class corresponds to the so called  $Y^{pq}$ quiver gauge theories.
This notation refers to a family of SE$_5$ manifolds. These manifold 
have an important application in the AdS/CFT correspondence, because they
describe the internal geometry probed by a stack of D3 branes 
probing the tip of the Calabi-Yau cones over the $Y^{pq}$ manifolds \cite{Benvenuti:2004dy}.

On the field theory side they correspond to quiver gauge theories, labeled by the positive
integer values of $p$ and $q$. For generic values of
$p$ and $q$ such theories have four abelian global symmetries. One of such symmetries is the
UV R-symmetry $R_0$, the others are two mesonic and one baryonic symmetries. 
In general the mesonic symmetry is $SU(2)\times U(1)$ and the baryonic symmetry is $U(1)$.
Further enhancement are possible for some values of $p$ and $q$.

%The partial topological twist breaks the non abelian symmetries and in the 2d theory one is left with an abelian symmetry that can potentially mix with the R-current.
%We will not discuss the second case, in which the extra baryonic symmetries are related to the non-anomalous $U(1)$s decoupling from the $U(N)$ gauge factors of the underlining gauge theories
%\footnote{In such cases the quiver becomes vector like and the geometries admit new deformations, i.e. new singularities far from the tip of the cone. In the KK reduction  
%there are new non-anomalous symmetries because there are new massless vectors in supergravity.}).

One of the main results of \cite{Benini:2015bwz} regards the role of the baryonic symmetries
in the compactification to two dimensions.
In the 4d case the exact R-current of the $Y^{pq}$ theories 
corresponds to a mixing of the abelian mesonic symmetry with $R_0$, while the coefficient of the mixing of  the baryonic symmetry vanishes.
Interestingly, by compactifying to 2d, the baryonic symmetry can 
contribute to the exact R-current, i.e. the mixing parameter of such baryonic symmetry 
is not necessarily  vanishing anymore.

One can see this mechanism at work explicitly. Let us turn the 
flavor symmetries off 
(i.e. we perform a topological twist just along the R-symmetry and the baryonic symmetry of the 4d theory).
The field strengths of the 4d theory can be written as 
\begin{equation}
\label{fluxes}
F_R^{(4D)} \rightarrow F_R^{(2D)} - \frac{\kappa}{2} t_\mathfrak{g}\,\,,
\quad\quad
F_{B}^{(4D)} \rightarrow F_{B}^{(2D)} + \epsilon_B  F_R^{(2D)}- B \, t_\mathfrak{g}\,\,,
\end{equation}
where  $\kappa$ is the normalized curvature of $\Sigma_\mathfrak{g}$, 
$t_\mathfrak{g}$ is the Chern class of the tangent bundle to $\Sigma_\mathfrak{g}$ normalized such that
$\int_{\Sigma_\mathfrak{g}} t_\mathfrak{g} = \eta_{\Sigma}$ \cite{Benini:2015bwz},
$B$ is  the magnetic flux for the weakly gauged $U(1)_{B}$ and
$\epsilon_B$ parameterizes the mixing of $U(1)_{B}$ with $U(1)_{R}$. 

The 2d central charge can be computed in terms of the 4d trace anomalies 
$k_{RRR}$ and $k_{RBB}$ (observe that $k_{BBB} = k_{RRB}=0$ as well as $k_R= k_B=0$ for $Y^{pq}$ theories).
This can be done by starting from the anomalous anomaly polynomial of the 4d theory
\begin{equation}
\label{I6}
I_6 = \frac{k_{RRR}}{6} c_1 (F_R)^3 +\frac{k_{RBB}}{6} c_1 (F_R) 
c_1 (F_B)^2 \,,
\end{equation}
where $c_1(F)$ is the first Chern class of the
bundle with curvature $F$.

Then one has to turn on the background fluxes (\ref{fluxes}) into (\ref{I6})
and integrate over $\Sigma_g$, obtaining the anomaly  polynomial
$I_4$ of the 2d theory.
The central charge $c_{2d}$ corresponds to $k_{RR}$, the coefficient  of the term $c_1(\mathcal{F}_R)^2$ in $I_4$.
Such central charge will be a function of the mixing coefficient $\epsilon_B$.
The exact central charge is obtained by extremizing $k_{RR}$ with respect to $\epsilon_B$.
\begin{figure}
\begin{center}
\includegraphics[width=10cm]{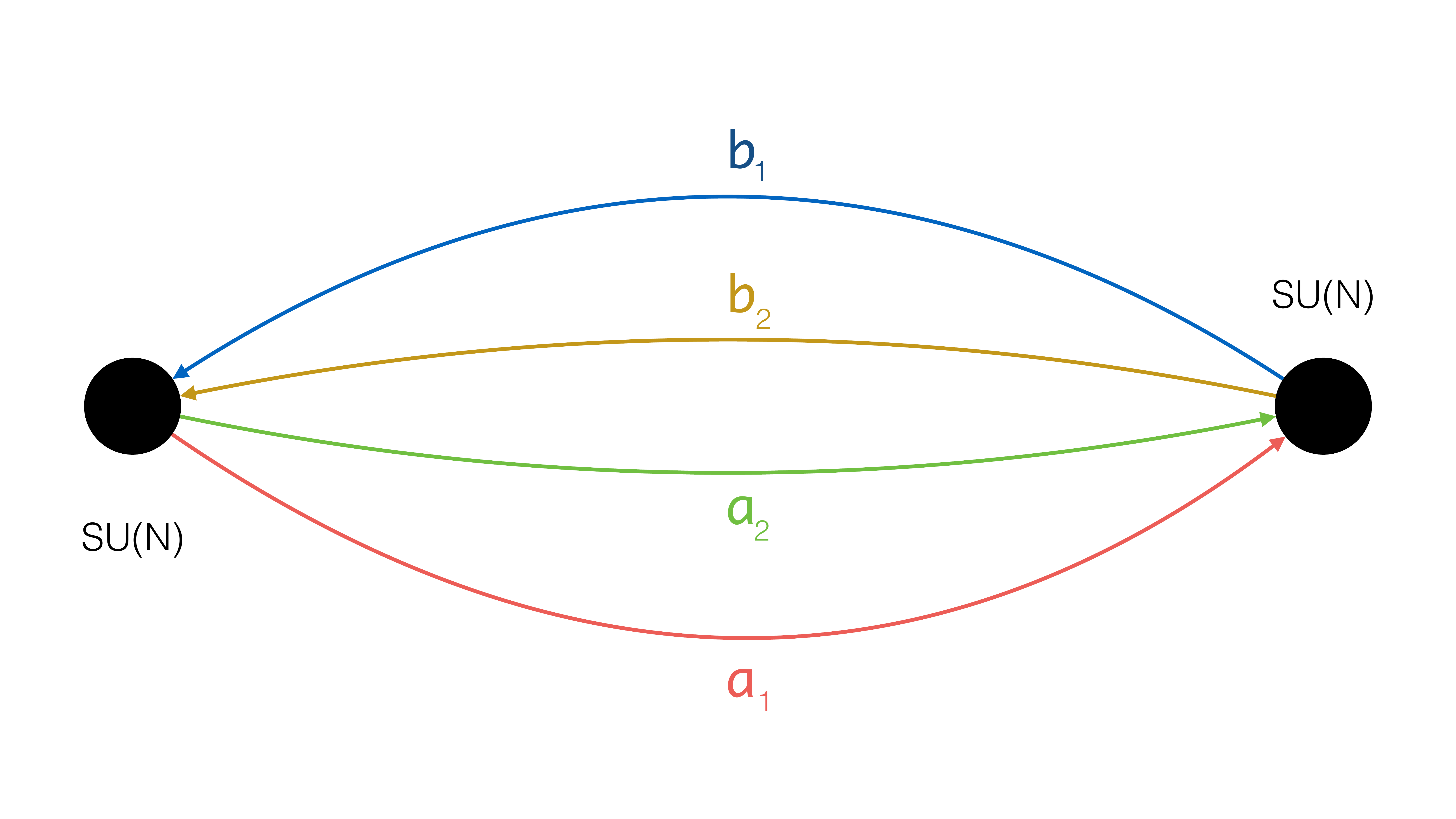}
\end{center}
\caption{Quiver gauge theory for the Klebanov-Witten theory, describing a stack of D3 branes probing the
$Y^{10}=T^{11}$ space.}
\label{KWquiver}
\end{figure}

Here we consider the case of
$p=1$ and $q=0$, corresponding to the Klebanov Witten theory \cite{Klebanov:1998hh}.
The associated quiver, represented in figure \ref{KWquiver}, has  two $SU(N)$ gauge groups connected by two pairs
of  chiral bifundamentals, $a_i$ and two anti-bifundamentals $b_i$, with $i=1,2$. 
This model has a superpotential 
$W = \epsilon_{ij} \epsilon_{lk} a_i b_l a_j b_k$, an R symmetry $U(1)_R$,
two $SU(2)$ flavor symmetries, rotating the matter fields and one baryonic symmetry,
assigning opposite charges to the fields $a_i$ and $b_i$.
At the fixed point the exact R-symmetry assigns charge $1/2$ to the chiral fields,
signaling the absence of mixing between the UV R-current and the other global symmetries.

This theory can be compactified on a Riemann surface along the lines of the discussion above,
by turning on a flux $\kappa$ for the R-current and $B$ the baryonic symmetry.
The central charge $c_{2d}$ can be obtained from the knowledge of the anomalies
$k_{RRR}$ and $k_{RBB}$.
By following the calculation explained above one finds that in this case
the mixing parameter is $\epsilon_B = 2B/\kappa$ and 
at large $N$ the central charge is
\begin{equation}
\label{ccFT}
c_{2d} = - \eta_{\Sigma} N^2
\left(\frac{9}{4}\kappa + \frac{12 B^2}{\kappa}\right)
\end{equation}
and a 2d SCFT can exist only for $\kappa=-1$.

As a last remark let us consider a twist of a 4d SCFT
along the 4d exact R-symmetry only.
In this case there is a general relation between  $a_{4d}$ and $c_{2d}$, first discussed in 
\cite{Johansen:2003hw} (see also \cite{Benini:2015bwz} for further discussions on this 
"universal twist"). Such relation is 
\begin{equation}
\label{FTca}
c_{2d} =  - \frac{3 \eta_{\Sigma} \kappa}{2}  k_{RRR} N^2 = 
-\frac{16 \eta_{\Sigma} \kappa }{3} a_{4d}\,,
\end{equation}
where, again, the last relation shows that $c_{2d}$ is positive  only for  $\kappa=-1$.

\section{Flowing from AdS$_5$ to AdS$_3$ in $\mathcal{N}=2$ gauged supergravity}
\label{sec:sugra35}

In this section we discuss some general aspects of the flow across dimensions
in gauged supergravity. We start from five dimensional $\mathcal{N}=2$ AdS$_5$ 
gauged supergravity with $n_V$  vector multiplets
(see appendix \ref{AppA}  for the details on five dimensional supergravity).
In this general discussion we do not consider the presence of hypermultiplets, because in some cases
they yield to massive vectors, i.e. to broken global symmetries in the SCFT.
We return on this problem at the end of the section.

The holographic dictionary identifies the $n_V+1$ vectors in the bulk 
with the R-current (the graviphoton) and  $n_V$ non-R global currents in the boundary.
These vectors can be thought as coming from a consistent truncation of 10D type IIB 
string theory on $AdS_5 \times Y_5$.
In such a case $n_V$ represents the sum of some of the internal (mesonic) symmetries of $Y_5$ and 
of some non anomalous baryonic symmetries, corresponding to the  
Betti  vectors in the  truncation. 
The global anomalies on the AdS$_5$ boundary theory are obtained by contracting the sections $X^I$ of the Special geometry  
with the coefficients of the CS terms in the action.
The holographic dictionary relates the central charge to the 
superpotential of the 5d theory 
 and $a$-maximization corresponds to the attractor mechanism for the
 scalars in the Special geometry \cite{Tachikawa:2005tq}.
  The mixing parameters correspond to the coefficients
 entering in the definition of the graviphoton \cite{Tachikawa:2005tq}.
  The other global anomalies are defined by a similar contraction
\cite{Barnes:2005bw}, involving the derivatives 
\footnote{This is because the terms $\partial_i X^I$ appear in $\delta_\epsilon \lambda$,
and the gauginos correspond to the fermionic components of the vector multiplets dual to the non-R global currents.}
of the sections $X^I$.

As mentioned already, the holographic RG flow connects AdS$_5$ vacua with AdS$_3 \times \Sigma_{\mathfrak{g}}$ geometries, where $\Sigma_{\mathfrak{g}}$
is a two dimensional Riemann surface with constant curvature and genus $\mathfrak{g}$. This is realized by for solutions of the form
  \begin{equation}\label{metric}
ds_5^2 = e^{2 f(r)} (-dt^2+dr^2+dz^2) + e^{2g(r)} d\sigma^2 \,,
\end{equation} 
with two radially dependent lapse functions, $f(r)$ and $g(r)$, and area element d$\sigma^2$ 
\be
d\sigma^2 = d\text{x}^2+ F(\text{x})^2 d\text{y}^2\,,
\ee
where $F(\text{x})$ is determined by the curvature $\kappa$ (or, alternatively, the genus $\mathfrak{g}$) of $\Sigma_{\mathfrak{g}}$
as follows
\begin{equation}
\label{acca}
\left\{
\begin{tabular}{llll}
$\mathfrak{g}=0$ & $\kappa=1$  & $F(\text{x}) = \sin (\text{x}) \qquad$ & $x\in [0,\pi]$, $y \in [0,2\pi[$\\
$\mathfrak{g}=1$ & $\kappa=0$  & $F(\text{x})=1 \qquad $ & $\text{x},\text{y} \in [0,1[$\\
$\mathfrak{g}>1$ & $\kappa=-1$ & $F(\text{x}) =\sinh(\text{x})$ & $x\in [0,\infty[$
\end{tabular}
\right.
\end{equation}
This metric ansatz encompasses both asymptotically locally AdS$_5$ spacetimes and solutions of the form AdS$_3 \times \Sigma_{\mathfrak{g}}$, hence the possibility of realizing the flow across dimensions.
 
The topological twist requires nonvanishing magnetic components of the the $n_V+1$ field strengths $F^I$
of the five dimensional theory. 
Here we consider a generic ansatz, corresponding to a topological twist along all of the 
$n_V+1$ abelian symmetries of the dual field theory 
\begin{equation}
\label{AnsatzRsugra}
F^I_{(5)} = -a^I vol_{\Sigma_{\mathfrak{g}}} + F^I_{(3)}
\end{equation}
where the volume form $vol_{\Sigma_{\mathfrak{g}}}$ is related to $\eta_{\Sigma}$ 
(introduced above) by $vol_{\Sigma_{\mathfrak{g}}} = 2 \pi \eta_{\Sigma}$ and 
$a^I$ are constant.

By working in units of $R_{AdS_5}=1$ 
the BPS equation relevant for obtaining the 2d central charge are
\footnote{Observe that in this paragraph we use the conventions of \cite{Maldacena:2000mw}.
We will modify the conventions in section \ref{sec:conifoldBetti} for making our notation
uniform with the literature.}
\begin{eqnarray}
f' +  e^f X^I P_I +\frac{e^{f-2g}}{2} a^I X_I  =0, \quad\quad
g' +  e^f X^I P_I - e^{f-2 g} a^I X_I  &=& 0 \,,
\end{eqnarray}
together with the constraint on the fluxes $3 P_I a^I = -\kappa$.
The AdS$_3$ fixed point is found by taking $f(r) = f_0 - \log r$ and constant $g= g^*$. In this way the two equations above reduce to a set of 
algebraic equations.
From such algebraic equation we can extract the relations
\begin{equation}
e^{f_0} = \frac{2 }{3(P^I X_I) }\,, 
\qquad
e^{2 g^*} = \frac{a^I X_I}{P^I X_I} \,,
\end{equation}
and 
the central charge is obtained from  the 
Brown-Henneaux formula \cite{Brown:1986nw}
\begin{equation}
\label{BrHe}
c_{2d} = \frac{3 R_{AdS_3}}{2 G_N^{(3)}} \,.
\end{equation}
The AdS$_3$ radius and the three dimensional Newton constant can be expressed in terms of 
higher dimensional quantities as
\begin{equation}
R_{AdS_3} = e^{f_0},
\quad 
G_{N}^{(3)} = \frac{G_N^{(5)}}{e^{2g^*} vol_{\Sigma_g}}
,
\quad 
\text{where}
\quad
G_N^{(5)} = \frac{vol(Y_5)}{2 \pi^2 N^2},
\quad 
vol_{\Sigma_g} = 2 \pi \eta_{\Sigma}\,.
\end{equation}
By using these relations we arrive at the following expression for the central charge 
computed from the supergravity data
\begin{equation}
\label{eq:csug}
c_{2d} =  \frac{6 \pi^3 N^2 \eta_{\Sigma}}{vol(Y_5)} e^{f_0+2g^*}
= \frac{4 \pi^3 N^2 \eta_{\Sigma}}{vol(Y_5) }
\frac{ a^I X_I}{(P^I X_I)^2 }\,.
\end{equation}
One can further refine such expression  by using 
the constraint on the Special geometry which relates
$X^I$ to $X_I$. In the normalization that we are considering here
the relation is 
\begin{equation}
\label{Xaltibassi}
X_I= \frac{1}{6} C_{IJK} X^J X^K\,.
\end{equation}

The R-charges, i.e. the coefficients of the 
mixing of the global currents in the expression of the R-current, can be read from the the graviphoton $A_\mu^I X_I$ \cite{deWit:2003ja}: the superconformal R-symmetry is read off from the anti-commutator of the supercharges acting on the scalar fields, as explained in \cite{Karndumri:2013iqa}\footnote{This observation was already made in \cite{Tachikawa:2005tq} and allowed the identification of the $U(1)$ R-symmetry in the 5d case.}. 
 The sections $X_I$ are proportional to the mixing coefficients, the R-charges, in the expression of the graviphoton,
$r^I = q X^I$.
The correct normalization $q$ on these coefficients can be found by fixing the charge of the gravitino to be one.
The gravitino charge is given by the expression $r^I P_I$ \cite{Tachikawa:2005tq}, and this fixes the proportionality constant to 
$t = 1/(X^J P_J)$.
%The normalization of the R-charge proceeds as in the 5d case,
%by fixing the charge of the gravitino $r^I P_I$ to unity. Eventually the R-charges are defined as
 %$r^I \equiv X^I/(X^J P_J)$.
By considering the relation (\ref{Xaltibassi})  and the definition of the R-charges,
the central charge becomes
\begin{equation}
\label{condep}
c_{2d} = \frac{2 \pi^3 N^2 \eta_{\Sigma}}{3 \, vol(Y_5)} C_{IJK} a^I r^J r^K\,.
\end{equation}

Some comments are in order.

Observe that the structure of the graviphoton exploits the relation between the R-charges 
on the field theory side and the R-charges on the gravitational dual description.
Consider a SCFT with $n_V$ global symmetries in addition to the R-symmetry.
In this case we can combine the R-current and the global currents to obtain 
$n_V+1$ trial R-currents, each one charging the superpotential, with non-vanishing R-charge.
Let us define these generators as $T_I$. The exact R-current in this case is a linear combination
of these trial R-currents, $T_R= \epsilon_I T_I$, with the constraint on the mixing 
parameters $\sum_I \epsilon_I=2$. 
The partial topological twist is performed along
the generator $T = n_I T_I$, with $\sum n_I=\kappa/2$.
With this choice we can map the generators $T_I$ with the vector multiplet gauge fields on the gravity side. The $R$-charges $\epsilon_I$ and the fluxes $n_I$ are proportional to the charges $r_I$ and the fluxes $a_I$
discussed on the gravitational side. We will see this mapping explicitly at work in the examples below.

The formulation of $c_{2d}$  in (\ref{condep}) is interesting because it is a quadratic expression in the  
R-charges, as expected from the field theory side. Moreover the extremization
corresponds to an attractor mechanism for the scalars
\footnote{See \cite{Karndumri:2013iqa,Karndumri:2013dca} for a complemetary discussion.}
of the Special geometry, in analogy 
with the case of $a$-maximization from  AdS$_5$ gauged supergravity discussed in \cite{Tachikawa:2005tq} and to the case of 3d $\tau$-minimization from AdS$_4$ gauged supergravity discussed in \cite{Amariti:2015ybz}.
This represents also the 5d dimensional version of the result of \cite{Benini:2015eyy}, where the flow from AdS$_4$ to AdS$_2$  in presence of fluxes
was studied. In that case the mixing of the graviphoton with the other vector multiplets
is related to an attractor mechanism and it is dual to the mixing of the
global currents in the exact R-current. This last can be obtained by extremizing the
topologically twisted index of \cite{Benini:2015noa}. 

The attractor mechanism can be understood as follows.
We have obtained the central charge from gauged supergravity 
parameterized in terms of a  combination of the sections, corresponding to the R-charges.
The value of the central charge at the fixed point
and the exact mixing are obtained from the analysis of the gaugino variations.
Indeed the vector multiplets are associated to the global symmetries  
mixing with the R-current at the fixed point.
At the supersymmetric vacuum the variations lead to the equations
(we will provide a detailed derivation in the appendix \ref{variation} for the case
of two vector multiplet, but the result is more general)
\begin{equation} \label{var_gaugino}
\partial_{x} (a^I X_I)-2 e^{2 g} \partial_{x} (P_I X^I) = 0 \,,
\end{equation}
where the index $x$ refers to the scalar $\phi^x$ in the Special geometry
(see appendix \ref{AppA} for review).
Now we consider the equation (\ref{eq:csug}),
compute the derivative with respect to the field $\phi^x$: by making use of \eqref{var_gaugino} we obtain

\begin{equation}
\partial_x c_{2d} \propto \frac{1}{(X^I P_I)^2}(\partial_{x} (a^I X_I)-2 e^{2 g} \partial_{x} (P_I X^I)) = 0 \,.
\end{equation}
In other words c-extremization corresponds, in gauged supergravity, to the
extremization of a superpotential function $W_3$ built out of the two quantities $X^I P_I$, corresponding
formally  to the superpotential 
of the 5d theory and $a^I X_I$, corresponding to the central charge of the gauging:
\be
\label{eq:W3}
W_3 \equiv 2 e^{-2g} \big(e^{-2 g} (X_I a^I)-2 (X^I P_I) \big)\,.
\ee

Lastly, in the case of the
"universal  twist" there is a relation between $c_{2d}$ and $a_{4d}$. In this case one considers a truncation with a graviton 
(see for example \cite{Buchel:2006gb} for generic truncations on $Y^{pq}$ manifolds) with
$X_0= 1/(3X^0)$, $P_0 =1/3$ and  $a^0 = -\kappa$.
By substituting in (\ref{eq:csug})
we have
\begin{equation} 
c_{2d} = \frac{4 \pi^3 N^2 \eta_{\Sigma}}{3 \, vol(Y_5)}  =
- \frac{16 \eta_{\Sigma} \kappa}{3} a_{4d},
\quad
\text{with}
\quad
a_{4d} =  \frac{\pi^3 N^2 }{4 \, vol(Y_5)} \,,
\end{equation}
that requires $\kappa=-1$ and corresponds to the field theory result of formula (\ref{FTca}).

In the following we apply this formalism  to some examples.
First we reproduce the results of  \cite{Benini:2013cda}, where $Y_5=S^5$.
Then we discuss the $T^{1,1}$ truncation. In this case we check the result by considering the twist along a combination 
of the 4d exact R-symmetry and the baryonic symmetry. As already mentioned, the  $T^{1,1}$  reduction yields a model of (abelian) $\mathcal{N}=2$ 5d gauged supergravity. This theory is characterized by the presence of a
Betti vector multiplet and hypermultiplets as well. The analysis in presence of hypermultiplets is more complicated and 
in the next paragraph we will discuss its modification.

\subsection{Inclusion of the hypermultiplets}
\label{hyperlagrange}

When considering the hypermultiplets $q^X$ there can be massive vector multiplets in the spectrum 
\footnote{Observe that it is not always the case, for example consistent truncations with Betti 
hypermultiplets turned do not higgs any further symmetries and do not lead to massive vectors.}.
On the supergravity side the condition for massive vectors is $K_I^X X^I=0$ (see for example\cite{Louis:2012ux}).
As reviewed in \cite{Tachikawa:2005tq},
by expanding the prepotentials around the origin one has, at linearized level, 
$P_I = P_{I(0)} + m_I^a P_a$, where $P_a$ plays the role of a Lagrange multiplier.
The holographic dictionary associates these massive vector bosons to 
anomalous global currents. Their contribution can be added to $a_{4d}$ by introducing the multiplier:
this sets a constraint preventing the massive vectors 
to mix with the graviphoton.

Turning on the fluxes and flowing to the AdS$_3$ vacuum in general triggers the flow of the hyperscalars as well.
Nevertheless at the AdS$_3$ fixed point the constraint on the massive
vector needs to be satisfied as well. 
By enforcing the constraint in the relation
between the sections in the AdS$_3$ vacuum 
we forbid the mixing of the massive vector with the 
graviphoton.
As we will see in detail later, at the AdS$_3$ point a 
further constraint on the fluxes  $a^I$ needs to hold. Such constraint, coming from the BPS equations, 
has to be imposed  to recover the structure of the topological twist.
Physically speaking this corresponds to the gravity dual mechanism of a consistency requirement for the topological twist, i.e. on the field theory side one can perform the
topological twist only along conserved non anomalous global abelian currents.

\section{$\mathcal{N}=4$ SYM and $U(1)^3$ gauged supergravity}
\label{sec:n4}
As a first example we  reproduce the results of  \cite{Benini:2013cda}, namely we 
compute the two dimensional central charge 
obtained by the R-symmetry twist of 4d $\mathcal{N}=4$ SYM
to 2d $\mathcal{N}=(0,2)$.
In this case there is an $SO(6)_{R}$ symmetry and the twist is done
along the $SO(2)_R^3$ abelian  subgroup, i.e.
$$
T = a^I T_I \,,
$$
where $T_I$ are the generators of the abelian symmetries and 
$a^I$ correspond to the magnetic fluxes of the weakly gauged abelian factors
\footnote{Observe that here we are keeping the same notations for the indices that we adopted on the supergravity side, even if there is no meaning 
in rising and lowering the index $I$ in field theory. Here the index $I$ runs over the Cartan subgroup of the global symmetry group.}.
If $\sum a^I = -\kappa$  and for generic $a^I$, supersymmetry is broken to $\mathcal{N}=(0,2)$ \cite{Benini:2013cda}.
The 2d R-current $T_R$ is a linear combination of the three abelian symmetries
$T_{R}^{(2d)} = \epsilon^I T_I$, where the constraint $\sum \epsilon^I = 2$ has to be enforced.
The central charge $c_{2d}$  has been computed in \cite{Benini:2015bwz} as a function of such mixing.
This has been extremized in terms of $\epsilon^I$.
At the fixed point such mixing parameters are
\begin{equation}
\epsilon^I = \frac{2 a^I(2a^I + \kappa)}{\theta}
\quad \text{with} \quad
\theta \equiv 2\sum_{I=1}^3 (a^I)^2 - \big(\sum_{I=1}^{3} a^I \big)^2
\end{equation}
and the central charge becomes 
\begin{equation}
\label{FT}
c_{2d} = -12 \eta_{\Sigma}N^2 \frac{a^0 a^1 a^2}{\theta}\,.
\end{equation}
In what follows we want to reproduce this result in the gauged
supergravity setup.
In this case $Y_5=S^5$ and we use the conventions of \cite{Benini:2013cda}, i.e.  \begin{equation}
R_{AdS_5}=1, \quad \quad P_I = \frac{1}{3}, \quad  \quad 3X_I = (X^I)^{-1} \,.
\end{equation}
The constraint on the fluxes is $ 3 a^I P_I= \sum a^I = -\kappa$ and
we define the R-charges as $r^I = X^I/(X^JP_J)$.
By plugging in these values in (\ref{condep}) the central charge is 
\begin{eqnarray}
c_{2d} = 
 \frac{4 }{3} N^2 \eta_{\Sigma} (a^0 r^1 r^2 + a^1 r^2 r^0 + a^2 r^0 r^1)\,.
\end{eqnarray}
We extremize this function with the constraint on the R-charges
$\sum r^I = 3$.
At the fixed point the charges are
\begin{eqnarray}
r^0 =  \frac{3 a^0}{\theta} \left(a^0-a^1-a^2\right)
,\quad \quad
r^1 = \frac{3 a^1}{\theta} \left(a^1-a^0-a^2\right) 
\end{eqnarray} 
and the central charge matches the field theory  result (\ref{FT}) for $\kappa=-1$.

\section{$T^{1,1}$  with a Betti multiplet}
\label{sec:conifoldBetti}
In this section we discuss the main result of our paper.
We study a consistent  $\mathcal{N}=2$ 
truncation of $T^{1,1}$ with a Betti multiplet \cite{Cassani:2010na}
and we reproduce the mixing of this vector with the graviphoton of the 
5d theory when flowing to AdS$_3$.
Let us first recall some general aspects of the baryonic symmetries and their relation with the Betti vector multiplets.
We keep the discussion general so far.
When considering the KK spectrum of IIB supergravity on AdS$_5 \times$SE$_5$
one can see the emergence of new massless matter arising from the non trivial second cohomology group of the SE$_5$ manifold.
In general a truncation to 5d $\mathcal{N}=4$ SUGRA yields b$_3$(SE$_5$) of such
vector multiplets, where $b_3$ is the third Betti number of SE$_5$.
Such $\mathcal{N}=4$ vector multiplets, called Betti vectors, can be further truncated to $\mathcal{N}=2$ vector multiplets or  $\mathcal{N}=2$ Betti hypermultiplets.
The former are dual to non anomalous conserved baryonic currents \cite{Ceresole:1999ht} (which are those we consider) while the latter are associated to exactly marginal operators.

We will be interested to truncation leaving  the Betti vector multiplets in the spectrum
\footnote{See \cite{Donos:2014eua} for a reduction to AdS$_3$ with the Betti hypermultiplet turned on.}.
Such a truncation has been explicitly obtained for the case of $T^{1,1}$ in \cite{Cassani:2010na}
(see also \cite{Halmagyi:2011yd}).
In this case b$_3(T^{1,1})=1$, i.e. there is a single Betti vector and it is dual to the baryonic symmetry of the Klebanov Witten theory.
On the field theory side this baryonic symmetry can be identified as follows. The model has two $U(N)$ gauge groups. In the IR the dynamical gauge groups is $SU(N)^2$, i.e. the two $U(1)$ symmetry become flavor symmetries, one of which is redundant. The leftover $U(1)$ corresponds to the baryonic symmetry of interest.
Such symmetry does not mix with the R-charge in 4d. This is a general statement, see for example \cite{Butti:2005vn} for a general discussion.
% that can be reformulated on the supergravity side by observing that the Betti vectors are singlets under the symmetries of the SE$_5$ manifold.

The $\mathcal{N}=2$ truncation of \cite{Cassani:2010na} has a third vector, which is Higgsed due to the presence of extra hypermultiplets, containing a 
Stuckelberg field\footnote{See for example \cite{Louis:2012ux} for a general discussion 
on such higgsing in gauged supergravity.}. 
This requires some modification to our general discussion, along the lines of section (\ref{hyperlagrange}).

The section is organized as follows. First we introduce the details of the truncation
in 5d, then we study the AdS$_3$ vacuum.
We perform the numeric analysis to study the holographic RG flow
from the AdS$_5$ UV fixed point to the AdS$_3$ IR one.
We conclude by computing $c_{2d}$ and the structure of the mixing of the global currents  
of the 2d theory with the formalism developed above,
matching the result with the one obtained in field theory. Let us finally mention that BPS solutions interpolating between an AdS$_4$ UV fixed point and AdS$_2$, in theories with hypermultiplets and Betti multiplets have recently been found in \cite{Halmagyi:2013sla}.

\subsection{The 5d truncation}

Here we review the truncation of 10d type IIB  on $ T^{1,1}$ leading to 5d gauged supergravity with a Betti vector, the so-called "Betti-vector" truncation of \cite{Cassani:2010na}, subsequently studied in \cite{Halmagyi:2011yd}. We refer the reader to these articles for further details. This model consists of the $\mathcal{N}=2$ graviton multiplet, two  
$\mathcal{N}=2$ vector multiplets and two hypermultiplets (one of which is the universal hypermultiplet), and it admits a $\mathcal{N}=2$ supersymmetric AdS$_5$ vacuum.

Schematically, the bosonic matter content is \cite{Halmagyi:2011yd}
\begin{itemize}
\item gravity +2 vectors: $(g_{\mu\nu}, A_0, A_1,A_2, u_2,u_3)$;
\item 2 hypers: $(u_1,k,\tau,\bar{\tau},b_0^i,\bar{b}_0^i), i=1,2$.
\end{itemize}

In this truncation one of the vectors becomes massive via Higgs mechanism, and this vector multiplet plus one hypermultiplet become a massive vector multiplet.  The matter content can also be rewritten in the following way (Table 2 of \cite{Cassani:2010na}):
\begin{itemize}
\item the gravity multiplet, containing a graviton $g_{\mu \nu}$ and a massless vector $A_g$;
\item A massless Betti vector multiplet, containing  a massless vector $A_{\mathcal{B}}$
and one real scalar $u_2$ with $m^2 =-4$ ($\Delta=2$);
\item A universal hypermultiplet, with 4 real fields $\tau, \bar{\tau}, b_0^1 , \bar{b}_0^1$, with $m_{\tau}^2=0$ ($\Delta=4$) 
and $m_{b_0}^2= -3$ ($\Delta=3$);
\item A massive vector multiplet, containing a massive vector $A_m$ (who has eaten its axion $k$) with
$m^2=24$, $\Delta =7$, and 4 real scalars $u_3, u_1, b_0^2, \bar{b}_0^2$ with $m^2= 12,21,21,32$ ($\Delta=6,7,7,8$). 
\end{itemize}
The quaternionic Killing vectors, specifying the gauging, are:
\begin{eqnarray}
K_0  = (3i b_0^i \partial_{b_0^i} +c.c.) -  2 \partial_k
,\quad
K_1 = 2 \partial_k 
,\quad
K_2 = 2 \partial_k
\end{eqnarray}
and the Killing prepotentials are 
\begin{eqnarray} \label{quat_prep}
P_0 & =& -i [(3 -\frac12 e^{-4 u_1} e^Z) \sigma_3 -2i e^{-2 u_1} v_i f_0^i \sigma_+ + 2i e^{-2 u_1} \bar{v}_i \bar{f}_0^i \sigma_-\,,  \nonumber \\
P_1 &=& -i e^{-4 u_1} \sigma_3\,, \nonumber \\
P_2 &=& -i e^{-4 u_1} \sigma_3\,,
\end{eqnarray}
where $v_i$ is the $SL(2,R)$ vielbein and
\be
f_0^i= 3 i b_0^i\,, \qquad
e^Z= 4 - \frac{2i}{3} \epsilon_{ij} (f_0^i \bar{f}_0^j - \bar{f}_0^i \,.f_0^j)\,.
\ee

We can moreover express the vectors $A^0, A^1,A^2$ appearing in the model in terms of the Betti vector $A_{\mathcal{B}}$, the massless vector appearing in SE$_5$ truncation $A_g$ and the massive vector $A_m$, in this way
\vspace{2mm}
%For $Q=4$ we have $ V = -12 \Lambda =-12$ hence $\Lambda =-1$.
\begin{eqnarray}\label{vectinbetti}
A_0 & = &\frac{1}{3} (-\sqrt{6} \, A_m + \sqrt{3} A_g )  \,, \nonumber \\
A_1 & = & \frac{1}{6 }  \left( \sqrt{6} \,  A_m - 3 \sqrt2 A_{\mathcal{B}} +  2 \sqrt{3}  A_g  \right) \,,\nonumber \\
A_2 &=& \frac{1}{6}  \left(  \sqrt{6} \,  A_m +3 \sqrt2   A_{\mathcal{B}} + 2 \sqrt{3} \, A_g \right) \,,
\end{eqnarray}
the inverse relations being
\be \label{decompose}
A_{\mathcal{B}}  = \frac{A_2 -A_1}{\sqrt2 } \,,  \qquad
A_m  =  \frac{A_1 +A_2- 2 A_0}{\sqrt{6}} \,, \qquad
A_g = \frac{A_0 +A_1 +A_2}{\sqrt{3}}\,.
\ee
One can see that when $A^2=A^1$ the Betti vector vanishes, and, upon a suitable truncation of the scalars, we are back to the SE$_5$ truncation. The expressions we just wrote will be useful when dealing with the explicit form of the mixing for the baryonic symmetries in the R-symmetry.

Let us finally mention that the AdS$_5$ vacuum is obtained for the following values of the scalar fields:
\be \label{vacuum1}
u_{1,AdS_5}=u_{2,AdS_5}=u_{3,AdS_5}=b_{0,AdS_5}^1=b_{0,AdS_5}^2=\tau=0\,,
\ee

\subsection{The ansatz and the BPS equations}

We study now the flow from AdS$_5$ to AdS$_3 \times \Sigma_g$ by solving the BPS equations. We consider the metric ansatz discussed  
 in section \ref{sec:sugra35}, which we rewrite here for convenience:
 \begin{equation}
ds_5^2 = e^{2 f(r)} (-dt^2+dr^2+dz^2) + e^{2g(r)} d\sigma^2 \,.
\end{equation} The gauge fields read
 \begin{equation}
\left\{
\begin{tabular}{lll}\label{gaugfield}
$\mathfrak{g}=0$ & $\kappa=1$  & $A_{\text{x}}^I = 0, \,\, A_{\text{y}}^I = a^I\cos(\text{x}),$\\
$\mathfrak{g}=1$ & $\kappa=0$  & $A_{\text{x}}^I =- \frac12 a^I \text{y}, \,\, A_{\text{y}}^I = \frac12 a^I \text{x},$\\
$\mathfrak{g}>1$ & $\kappa=-1$ & $A_{\text{x}}^I = 0, \,\, A_{\text{y}}^I = - a^I \cosh(\text{x}),$
\end{tabular}
\right.
\end{equation}
hence the field strengths, defined as  $F_{\mu \nu}^I = \partial_{\mu} A_{\nu}^I  - \partial_{\nu} A_{\mu}^I $ read
\be
F^I= -a^I F(\text{x})\,d\text{x} \wedge d\text{y} \qquad I= 0,1,2\,.
\ee
We will be interested here in the purely magnetic case\footnote{$t$ and $z$ components of the gauge fields of the form $A_t^I= q(r)$ and
$
A^I_z= w^I=const
$
are compatible with the Maxwell and Bianchi equations as well, and contribute to the black string electric charge $\int_{\Sigma_g} G_{I}$,
where $G_{I, \mu \nu} = {\epsilon_{\mu \nu}}^{ \sigma \rho }\frac{\partial \mathcal{L}} { \partial F_I^{\sigma \rho}}$. We will not consider them here.}.
The scalar fields of the Special geometry are positive, satisfy $X^0 X^1 X^2 =1$ and they are parameterized as
\be
\label{eq:X}
X^0= e^{4 u_3} \,, \qquad X^1=  e^{2u_2 -2u_3}\,, \qquad X^2= e^{   -2u_2 -2u_3 }  \,, \qquad X_I =  (X^I)^{-1}\,.
\ee
The two real scalars $u_2$ and $u_3$ are  independent  and belong to  the vector multiplets. 

We will make one simplifying assumption while looking for solutions of the BPS equations. First of all, as shown in \cite{Cassani:2010na}, one can consistently truncate away the universal hypermultiplet fields. Hence $\tau, b_0^1$ will be set to constant values.
Moreover, the remaining fields $b_0^2$ appears quadratically in the action, hence can be consistently se to zero. This setup was shown to be a consistent truncation in \cite{Herzog:2009gd} as well, the field $u_1$ being called the resolution mode of 
$T^{1,1}$
\footnote{Our Lagrangian is the same as (7.7) of \cite{Herzog:2009gd} upon the following field identification  
\be
\eta= \frac{2(2u_1+3u_3)}{5} \,, \qquad \chi= \frac25 (3u_1-3u_3)\,, \qquad \lambda = 2u_2\,.
\ee}.

We will then look for explicit solutions where all the scalar fields  except $u_1,u_2,u_3$ are consistently set to zero. One can see from eq. \eqref{quat_prep} that this has the net effect of setting the components $r=1,2$ of the  
Quaternionic Killing prepotentials $P_{\Lambda}^r$ to zero:
\be
P_{\Lambda}^1=P_{\Lambda}^2=0\,,
\ee 
namely in our solutions, similarly to  \cite{Halmagyi:2013sla}, the prepotentials are aligned in the $\sigma^3 $ direction, namely $P_{I}^3 \equiv -i \,P_I \neq 0$. Moreover, this has the net effect of setting $\omega_{\mu i}^j=0$.

The full analysis of the supersymmetry variations is in Appendix \ref{variation}, where the BPS equations are obtained. In the case under consideration $Q^r=(0,0-1)$, hence the projection on the Killing spinor simplifies to
\be \label{proj_simpl}
\gamma_2  \epsilon_i =  {\sigma^3}_{ij} \epsilon^j \,, \qquad \gamma_{34} \epsilon_i = i  {\sigma^3}_{ij} \epsilon^j  \,, 
\ee
and the BPS equations then reduce to:
\begin{eqnarray}
f'+ \frac{1}3 e^{f} X^I P_I + \frac{1}{6}a^I X_I e^{-2g+f}&=&0  \nonumber \\
g' + \frac{1}3 e^f X^I P_I -\frac{1}{3} a^{I } X_I e^{-2g+f}  &=&0  \nonumber \\ 
 8 {u^2}' \,  
+\frac{ e^{-2g+f}}{2} a^I \partial_2 X_I    - e^f \partial_{2} X^I P_I & = & 0  \nonumber \\
 24 {u^3}' \,  
+\frac{ e^{-2g+f}}{2} a^I \partial_3 X_I    - e^f  \partial_{3} X^I P_I & = & 0  \nonumber \\
16 u_1' - 2 \, \partial_X (X^I P_I)& = & 0  \nonumber \\
 a^I K_I^X& =& 0 \label{ntlquant} \nonumber \\
a^I P_I  + \kappa & = & 0 \label{lastquant} \label{BPSeqn}\,,
\end{eqnarray}
where we used the fact that $g_{22}=8$, $g_{33} =24$, $g_{23}=0$ and the metric of the hypermultiplet $u_1$ is $g_{11} =16$. 
Lastly, one can see that the BPS equations \eqref{ntlquant}-\eqref{lastquant} select a particular value for the Dirac quantization conditions of the theory, 
correspondingly, on the field theory side, to the
quantization of the fluxes through the  Riemann surface.

\subsection{The AdS$_3$ vacuum}\label{ads3vacuumsect}

In order for the metric to approach the AdS$_3 \times \Sigma_{\mathfrak{g}}$ fixed point, the metric functions should attain the constant value $f(r)= f_0 - \log r$ and  $g = g*$.  The scalars as well assume constant values, which we denote by $\phi_X^*$ and $q_i^*$. 
At the AdS$_3$ vacuum the equations \eqref{BPSeqn} reduce to a set of  algebraic equations for $g^*$, $f_0$ and the scalars
\begin{eqnarray}
\label{BPSvac}
2 e^{2 g^*} (X^I P_I) &=& C_{IJK} a^I X^J X^K \nonumber \\
e^{f_0} (X^I P_I) &=& 2 \nonumber \\
2 e^{2 g^*} (X^1 P_1 - X^2 P_2) &=&X^0(a^2 X^1 - a^1 X^2)
 \nonumber \\
2 e^{2 g^*} (2X^0 P_0-X^1 P_1 - X^2 P_2) &=& a^1 X^0 X^2 + a^2 X^0 X^1-2a^0 X^1 X^2
\end{eqnarray}
with the constraints 
\begin{equation}
\label{constraints}
a^0 = -\kappa/3, \quad \quad
a^1+ a^2 = 2 a^0, \quad \quad
X^1+X^2 = 2 X^0\,.
\end{equation}
Notice that in order to obtain \eqref{BPSvac} and \eqref{constraints} the first one of the projections \eqref{proj_simpl} becomes unnecessary. In other words, while the entire flow preserves one quarter of supersymmetries, the supersymmetry is actually enhanced to 1/2 BPS at the near-horizon AdS$_3 \times \Sigma_{\mathfrak{g}}$  geometry. 

Let's now analyze the equations \eqref{BPSvac} and \eqref{constraints} obtained. The first two relations reproduce, in gauged supergravity,
the structure of fluxes turned on when performing the topological twist
on the field theory side. 
Here we parameterize the fluxes $a^1$ and $a^2$ as
\begin{equation}
a^1 =- \frac{1}{3} (\kappa - 4 b), \quad \quad
a^2 =- \frac{1}{3} (\kappa + 4 b)\,.
\end{equation}
The last constraint in (\ref{constraints}), corresponding to $K_X^I X_I=0$, 
signals the presence of a massive vector multiplet in the spectrum,
i.e. a non conserved abelian current on the field theory side.
This is consistent with the fact that there are only two independent conserved
charges $a^I$.
In the case of $\kappa=-1$ the AdS$_3$ vacuum is found at
\begin{eqnarray}
\label{ads3vac}
&&
e^{4 u_1} = \frac{3 \kappa^2+16 b^2}{3 \kappa^2},
\quad\quad
e^{4 u_2} =\frac{\kappa+4 b}{\kappa-4 b},
\quad\quad
e^{12 u_3}= \frac{\kappa^2}{\kappa^2-16 b^2},
\nonumber \\
&&
\quad\quad
e^{6g^*} =\frac{\left(16 b^2+3 \kappa^2\right)^3}{729\kappa \left(16 b^2-\kappa^2\right)},
\quad\quad
e^{3f_0}=\frac{8}{27} \frac{\kappa^2-16 b^2}{\kappa^2}\,.
\end{eqnarray}
At this point one is ready to compute the central charge, by plugging in the values of 
$g^*$ and $f_0$ in (\ref{ads3vac}) in the Brown-Henneaux formula \eqref{BrHe}:
\be
c_{2d} = -\eta_{\Sigma} N^2 \left(\frac{9 \kappa }{4}  +  \frac{ 12 b^2}{\kappa} \right)\,.
\ee
This shows that only solutions with $\kappa=-1$ are admitted, as expected from the field theory side.
Indeed the case $\kappa=0$ is singular, while the case $\kappa=1$ gives a negative 
central charge. This match with the central charge obtained on the field theory side \eqref{ccFT}.

\subsection{Holographic RG flow}
Here we numerically solve the BPS flow
equations, studying the holographic RG flow 
connecting the AdS$_5$ UV
vacuum and the AdS$_3$ IR one.
In order to simplify the numerical problem we rewrite 
the BPS equations in an easier and more compact form 
by a change of variables.

\begin{figure}
\begin{center}
\begin{tabular}{cc}
\includegraphics[width=7cm]{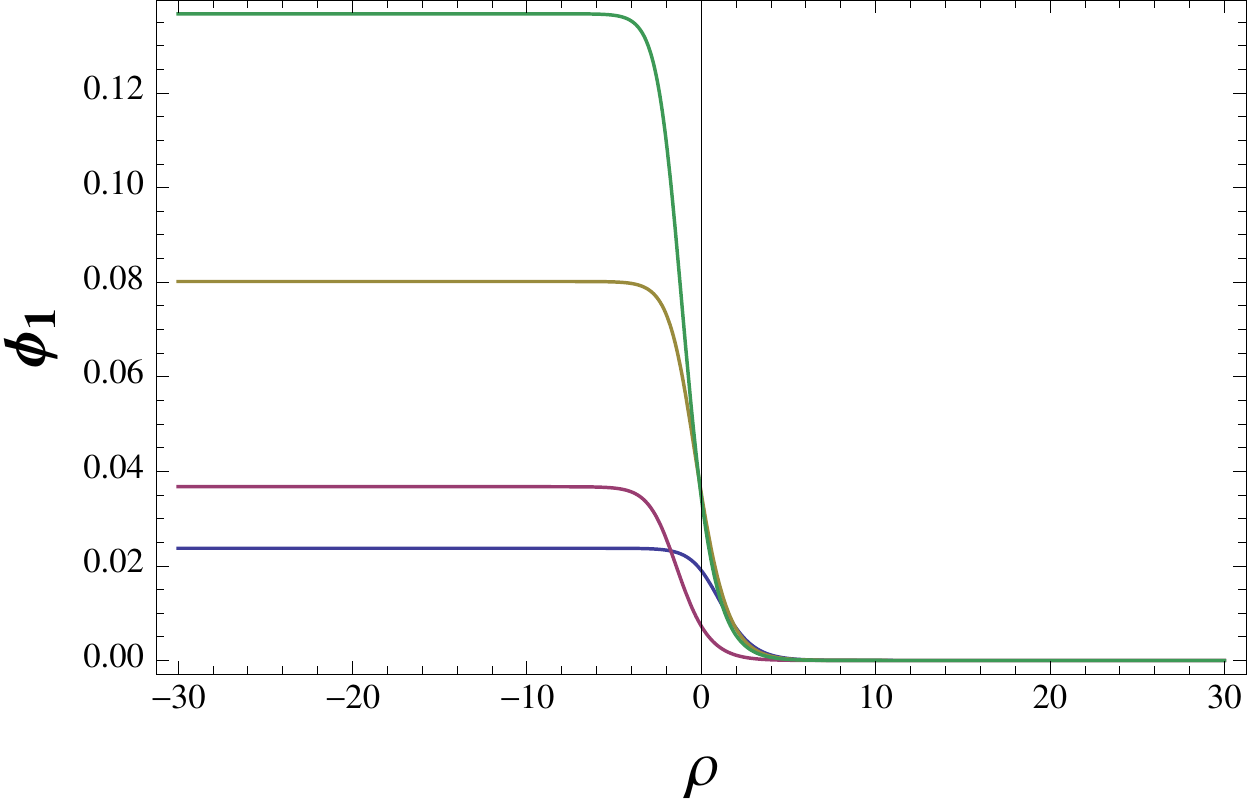}
&
\includegraphics[width=7cm]{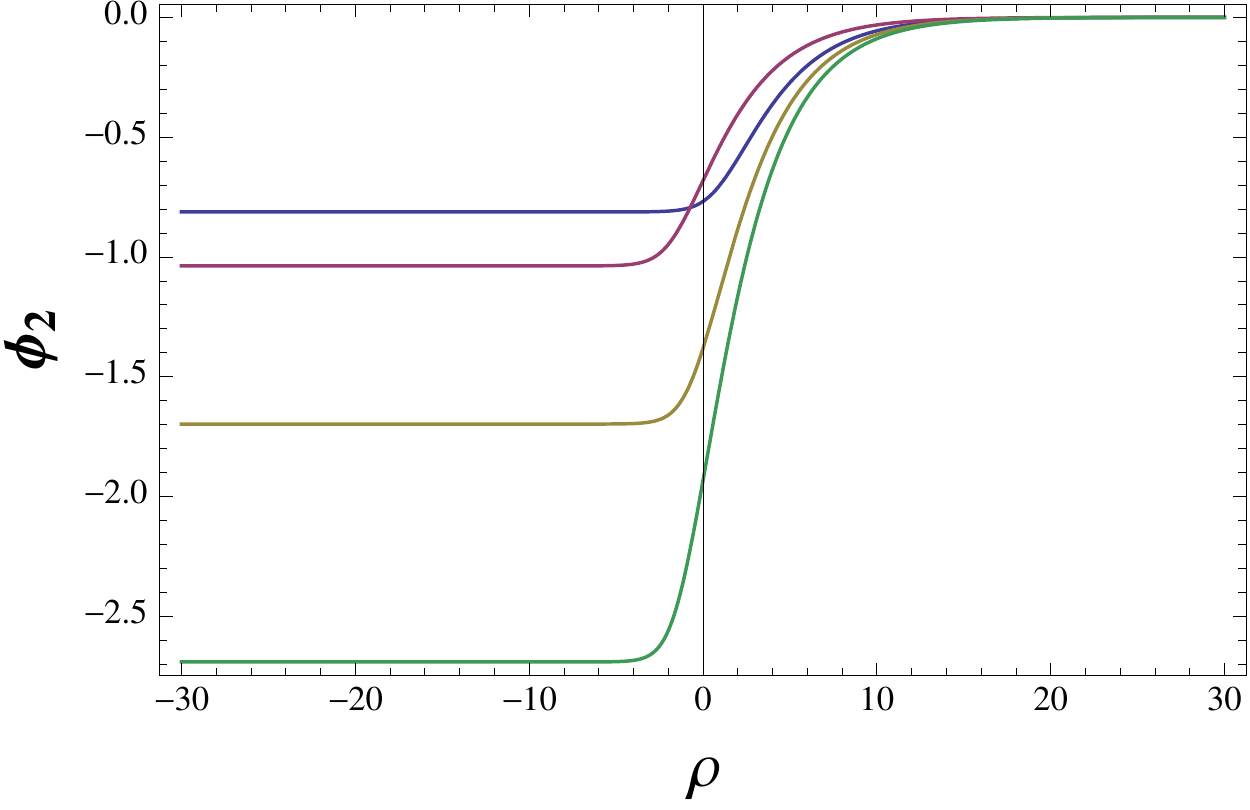}
\\
\includegraphics[width=7cm]{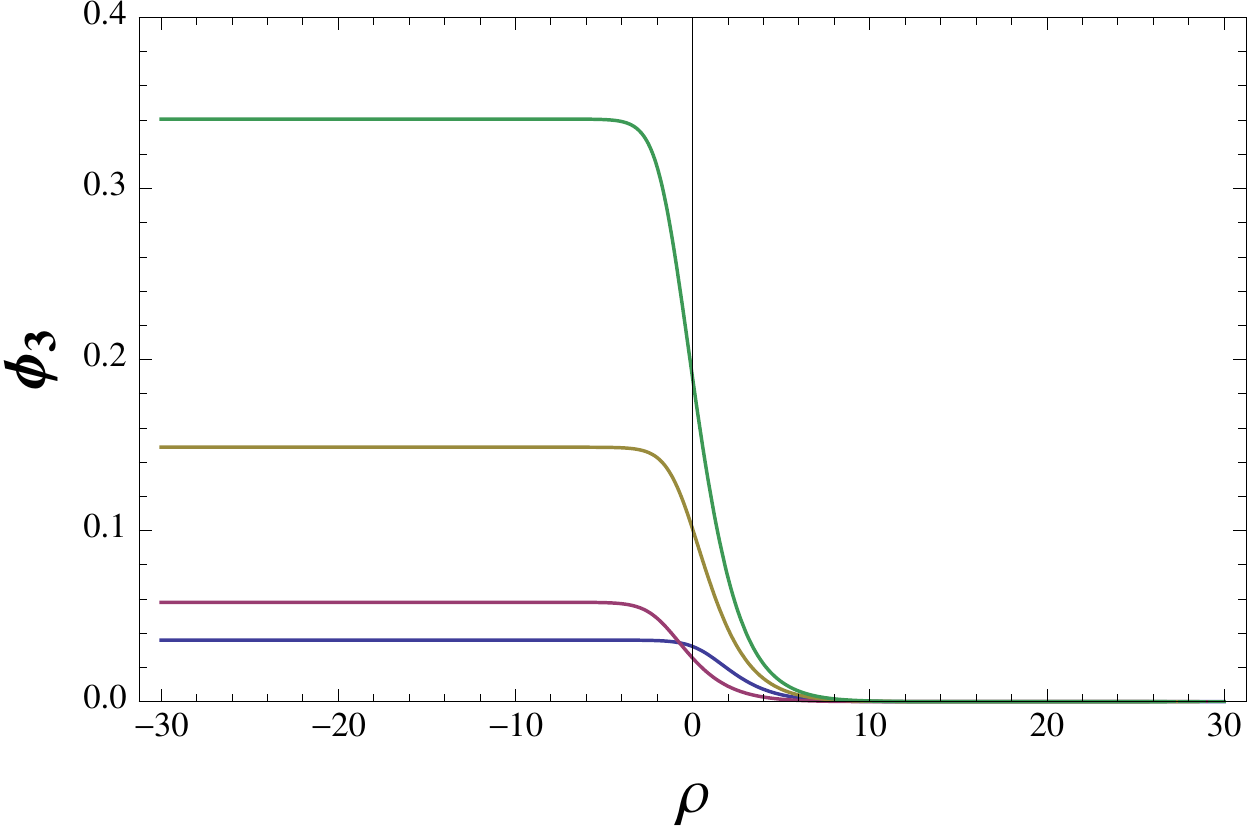}
&
\includegraphics[width=7cm]{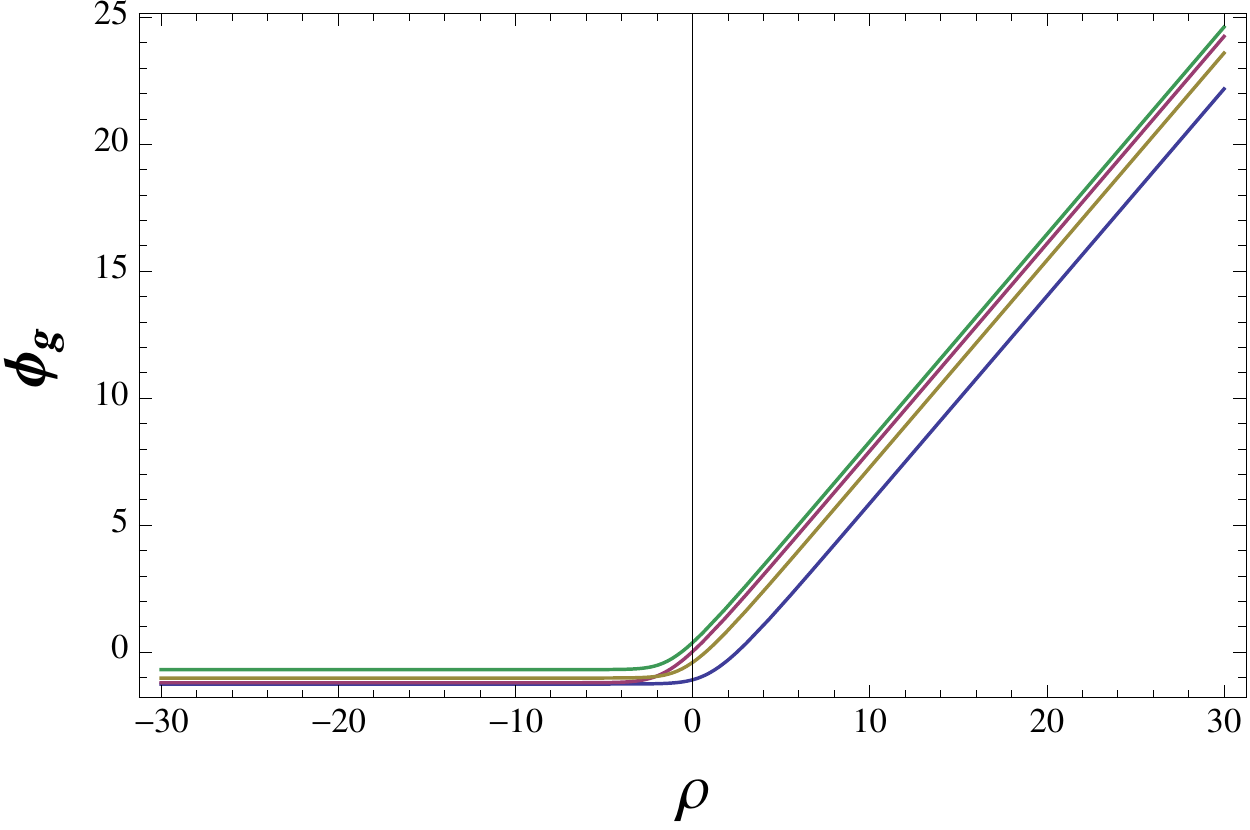}
\end{tabular}
\end{center}
\caption{Holographic RG flow for the of the scalar fields $\phi_i$ interpolating between the AdS$_5$ vacuum in the UV, $\rho \rightarrow \infty$ and the AdS$_3$ vacuum
in the IR, $\rho \rightarrow -\infty$.
The colors refers to the values $b=0.08$ (blue), $b=0.1$ (red), $b=0.15$ (yellow) and $b=0.2$ (green).}
\label{holoflow}
\end{figure}
This change of variables is analogous to the one discussed in 
\cite{Benini:2013cda,Bobev:2014jva}.
First one start by redefining the  radial coordinate as
\begin{equation}
\rho \equiv f+2g \,.
\end{equation}
The second step consists in considering a superpotential function
\footnote{This is the same superpotential that we already encountered in section \ref{sec:sugra35}
when we observed that c-extremization corresponds to an attractor mechanism in gauged supergravity.
This is possible because, even if we are in presence of hypermultiplets, we can still rotate the killing potentials
$P_I$ along the same direction.}
 for the theory on 
AdS$_3$. This corresponds to a combination of the AdS$_5$  superpotential $ X^I P_I$
discussed in \cite{Halmagyi:2011yd}  and of the central charge of the gauging 
 $ X_I a^I$. Their combination give rise to the 3d
superpotential function
(\ref{eq:W3}), that can be used to 
to reformulate the BPS equation in a compact
form.
The BPS equations for $f'$ and $g'$ can be recombined in
\begin{equation}
\label{rhor}
\frac{\partial \rho(r)}{\partial r} = \frac{e^{ \rho}}{2} W_3\,.
\end{equation}
The other BPS equations can be written as
\begin{equation}
\label{BPsimpler}
\frac{\partial{u_i}}{\partial r}= -\frac{1}{2 }e^{\rho} g^{ii} \partial_i W_3\,,
\quad \quad\quad
\frac{ \partial g(r)}{\partial{r}} =-\frac{e^\rho}{12} \frac{ \partial W_3}{\partial{g}}\,.
\end{equation}
These equations are more suitable for the numerical analysis, and they are independent of
$f$. Once we have the solution for these equations the function $f$ is determined automatically
from (\ref{rhor}).
Moreover, by using (\ref{rhor}),  the equations (\ref{BPsimpler}) become
\begin{equation}
\frac{\partial{u_i}}{\partial \rho}  = 
- g^{ii} \frac{1}{W_3} \frac{\partial W_3}{\partial u_i}\,,
\quad \quad\quad
\frac{\partial{g}}{\partial \rho}  = 
- \frac{1}{6 W_3} \frac{\partial W_3}{\partial g}\,.
\end{equation}
It follows that if we canonically normalize the fields $u_i$  by $\sqrt g^{ii}$
and $g$ by $\sqrt 6$, and define a new set of 
variables $\phi_1,\phi_2,\phi_3$ and $\phi_g$ the BPS equations can be written together as 
\begin{equation}
\frac{\partial{\phi_i}}{\partial \rho}  = 
- \frac{1}{ W_3} \frac{\partial W_3}{\partial \phi_i}\,.
\end{equation}
This formulation shows that a slight simplification appears when working in 3d versus 5d, because the 3d theory appears to be agnostic of the distinction between scalars from hypermultiplets and vector multiplets. This fact was noticed in \cite{Karndumri:2015sia} for the model treated in Sec. \ref{M5} (see related discussion in that section). Indeed, one sees that the BPS flow equations are generated by varying a \textit{single} superpotential function, which in the present case is denoted by $W_3$.

Coming back to our analysis, in this basis the AdS$_3$ vacuum corresponds to
\begin{eqnarray}
\label{ads3vacuum}
\phi_1 &=& \sqrt{\frac{1}{2}} \log \left(\frac{16 b^2+3 \kappa^2}{3 \kappa ^2}\right)\,,
\quad\quad
\phi_2 = \sqrt \frac{3}{2} \log \left(\frac{\kappa+4b}{\kappa -4 b}\right)\,,
\nonumber \\
\phi_3 &=&  \frac{1}{3} \log \left(\frac{\kappa^2}{\kappa ^2-16 b^2}\right)\,,
\quad\quad
\phi_g =\frac{1}{\sqrt{6}}\log \left(\frac{\left(16 b^2+3 \kappa ^2\right)^3}{729 \kappa  \left(16 b^2-\kappa ^2\right)}\right)\,.
\end{eqnarray}
We can now numerical solve these equations by fixing $\kappa=-1$ and by imposing the boundary conditions 
in the IR, i.e. the fact the the solution flows to the AdS$_3$ vacuum in formula (\ref{ads3vacuum}).
From the figure \ref{holoflow} we can observe that at large $\rho$ this theory flows 
to a different fixed point, in which the fields $\phi_i$ sit at the AdS$_5$ vacuum \eqref{vacuum1}.

\subsection{$c$-extremization from gauged supergravity}

In this section we are interested in the structure of the mixing of the abelian 
currents leading to $c_{2d}$ as a result of the extremization
procedure.
We study the $c$-extremization in gauged supergravity
along the lines of section \ref{sec:sugra35}.
Interestingly we see that the effect of the hypermultiplets can be easily included here
by enforcing the constraint on the massive vector.
This can be done by parametrizing the mixing in terms of the R-charges $r^I$ 
and imposing the constraint $2 X_0 = X_1+X_2$. In this way the R-charges become
\begin{equation}
r^0 = \frac{1}{3}, \quad \quad
r^1 = \frac{X^1}{3X^0} = \frac{1}{3}(1-\epsilon)\,, \quad \quad
r^2 = \frac{X^2}{3X^0} = \frac{2X^0-X^1}{3X^0} = \frac{1}{3}(1+\epsilon)\,.
\end{equation}
where we defined $\epsilon = - (X^1-X^0)/X^0$.
We reproduce 
the topological twist by identifying the 
R-symmetry and the baryonic symmetry 
with the graviphoton and with the Betti vector respectively.
Explicitly
\begin{equation}
a^I T_I = \frac{k}{3} (T_0 + T_1 +T_2) + \frac{4}{3} b(T_2-T_1)\,,
\end{equation}
hence the identifications between the field theory generators $T_R^{f.t.} $ and $T_B^{f.t.} $ and the
supergravity ones
\begin{equation}
T_R^{f.t.} =  \frac{3}{2} (T_0 + T_1 +T_2)
,\quad \quad
T_B^{f.t.} =  \frac{3}{4} (T_2-T_1)\,.
\end{equation}
This decomposition reflects eqn. \eqref{decompose}. By using the relation  $vol(T^{1,1}) = \frac{16 \pi^3}{27}$
the central charge can be written as
\footnote{Observe that in our conventions $2X_I = C_{IJK} X^J X^K$
and $e^f = 2/(X^I P_I)$.}
\begin{equation}
c_{2d} = 
 \frac{6 \pi^3 N^2 \eta_{\Sigma}}{ \, vol(T^{1,1})} C_{IJK} a^I r^J r^K
 = \frac{3 N^2 \eta_{\Sigma}}{4} (8 b \epsilon-\kappa (3-\epsilon^2) )\,.
\end{equation}
At the extremum, corresponding to  $\epsilon^* = -4b/\kappa$, the central charge becomes
\begin{equation}
c_{2d} = -\eta_{\Sigma} N^2 \left(\frac{9 \kappa }{4}  +  \frac{ 12 b^2}{\kappa} \right)\,,
\end{equation}
matching the field theory result (\ref{ccFT}) for $\kappa=-1$.
Observe that as a consistency check we can plug in $c_{2d}$ the values of the scalars at the vacuum
(\ref{ads3vac}) and obtain the value of $\epsilon^*$.

\section{M5 branes}
\label{M5}
In this section we discuss the flow to AdS$_3$ for a 5d theory constructed in \cite{Szepietowski:2012tb} from 
a consistent  truncation  of 7d
$\mathcal{N} = 4$ SO(5) gauged supergravity reduced on a Riemann surface.
The holographic 4d theories associated to such truncation were constructed in \cite{Bah:2012dg}.
They are quiver gauge theories, obtained from the class S theories of \cite{Gaiotto:2009we} by gluing the pants 
such that only $\mathcal{N}=1$ supersymmetry is preserved.
We will not review the construction here, and refer the reader to the original paper for details\footnote{This model was discussed in \cite{Karndumri:2015sia}. In the same paper the uplift in 10d is constructed and the authors showed that the R-symmetry agrees with the canonical R-symmetry in 11D based on the existing classification, \cite{Figueras:2007cn}.}.
We do not discuss the calculation of the central charge from the field theory data (see \cite{Baggio:2014hua}) and the 
analysis of the flow in gauged supergravity (see \cite{Benini:2013cda,Baggio:2014hua}).
Here we are just interested in constructing $c_{2d}$ by applying our formalism.
We can also check our result against the one obtained in \cite{Benini:2013cda}, where the reduction of the 
7d theory on the product of two Riemann surfaces $\Sigma_1 \times \Sigma_2$ was discussed.

By using our formalism, $c_{2d}$ can be obtained from 5d $\mathcal{N}=2$ gauged supergravity in terms of the 
prepotentials $P_I$.
 The supergravity analysis leads to the following Killing prepotentials\footnote{We are modifying 
the notations of \cite{Szepietowski:2012tb}, with the change of variables
$
\lambda _1=\frac{-4 u_1+u_2-u_3}{10}, 
\lambda _2=\frac{u_1-4 u_2+u_3}{10},
B=\frac{3(u_1+u_2-u_3)}{10} 
$. 
}
\begin{equation}
P_0 = e^{u_3}, \quad\quad
P_1 = 2+e^{u_3}\frac{z_1+\kappa_1}{8}
, \quad\quad
P_2 = 2-e^{u_3}\frac{z_1-\kappa_1}{8}\,.
\end{equation}
The sections are parameterized as
\begin{equation}
X^0 = e^{u_1+u_2},\quad \quad
X^1 = e^{-u_1},\quad\quad
X^2 = e^{-u_2}.
\end{equation}
By following the procedure discussed above we can reduce this theory on $\Sigma_2$,
with curvature $\kappa_2$.
The constraints on the fluxes are 
\begin{equation}
2 a^1 +2 a^2 = -\kappa_2
\quad\quad
8 a^0 +a^1 (\kappa_1-z_1) + a^2 (\kappa_1+z_1)=0\,.
\end{equation}
We can parameterize $a^1$ and $a^2$ as
\begin{equation}
a^1 = -\frac{\kappa_2-z_2}{4}\,,
\qquad
a^2 = -\frac{\kappa_2+z_2}{4}\,,
\end{equation}
where $z_2$ plays the role of the baryonic symmetry discussed in the case of $T^{1,1}$, i.e. it is associated to  
a massless vector multiplet that does not mix with the 5d graviphoton, but that can mix in 3d.
One can observe that $a^1+a^2 = -\kappa_2/2$ and 
$a^1-a^2 = z_2/2$ are the fluxes for the 5d graviphoton, that can be obtained from $a$-maximization,
and for the second vector multiplet, that couples in the two dimensional CFT as expected.
In this case there is a massive vector multiplet as well. Its effect can be understood from the 
hyperino variation, constraining the scalars as
\begin{equation}
8 e^{2(u_1+u_2)} + e^{u_1} (\kappa_1-z_1) + e^{u_2} (\kappa_1+z_1) = 0\,.
\end{equation}
Observe that the parametrization in terms of the scalars is unnecessary, and we
could simply provide the relation in terms of the constraint on the sections, 
$8 X^0 +X^1 (\kappa_1-z_1) + X^2 (\kappa_1+z_1)=0$.
This constraint allows us to define the $R$ charges $r^I = X^I/(X^IP_I)$ as
\begin{equation}
r^0 =\frac{4 z_1 \epsilon-\kappa_1}{16} \,,
\quad \quad
r^1=\frac{1}{4}-\epsilon \,,
\quad \quad
r^2=\frac{1}{4}+\epsilon\,.
\end{equation}
The central charge is proportional to $C_{IJK} a^I r^J r^K$, that in this case becomes
\begin{equation}
 C_{IJK} a^I r^J r^K=\frac{
 \kappa _1 \kappa _2 (3-16 \epsilon ^2) -8 \epsilon(\kappa _1 z_2+\kappa _2 z_1)-z_1 z_2 (1-48 \epsilon ^2) }{128}\,,
\end{equation}
which is extremized at
\begin{equation}
\epsilon =\frac{\kappa _1 z_2+\kappa _2 z_1}{4 \left(3 z_1 z_2-\kappa _1 \kappa _2\right)}\,.
\end{equation}
The overall factor in the calculation of the central charge can be
obtained by expressing $G_N^{(3)}$  in terms of $G_N^{(11)}$ . 
We start by providing the relations \cite{Henningson:1998gx}
\begin{equation}
G_N^{(11)}  = \frac{ \pi^4}{32 N^3}
,\quad \quad
G_N^{(7)}= \frac{G_N^{(11)}}{Vol_{S^4}}=\frac{3  \pi^2}{16 N^3}\,.
\end{equation}
The central charge can then be expressed as
\begin{eqnarray}
c_{2d} = \frac{3 R_{AdS_3}}{2 G_N^{(3)}} = \frac{3}{2} e^{f+2g} \frac{4 \pi^2 \eta_{\Sigma_1} \eta_{\Sigma_2}}{G_N^{(7)}}
=  8 N^3 \eta_{\Sigma_1} \eta_{\Sigma_2}C_{IJK} a^I r^{* J} r^{*K}\,.
\nonumber
\end{eqnarray}
We further need to normalize the fluxes as done in \cite{Benini:2013cda}. It corresponds
to the relation $\kappa_i \rightarrow 2\kappa_i$ and $z_i \rightarrow 2z_i$.
The final result is
\begin{equation}
c_{2d} = 
\frac{\eta _1 \eta _2 N^3 \left(3 \kappa _1^2 \kappa _2^2+\kappa _1^2 z_2^2-8 \kappa _1 \kappa _2 z_1 z_2+\kappa _2^2 z_1^2+3 z_1^2 z_2^2\right)}
{4 \left(\kappa _1 \kappa _2-3 z_1 z_2\right)}\,,
\end{equation}
that matches the one found in \cite{Benini:2013cda} at large $N$.
The finite $N$ contribution has been obtained in \cite{Baggio:2014hua} and involves higher derivatives in AdS$_5$
supergravity.

\section{Conclusions}
\label{sec:Conclusions}
In this paper we discussed some aspects of supersymmetric flows from
AdS$_5$ to AdS$_3 \times \Sigma_g$ in gauged supergravity, dual to the compactification of 4d $\mathcal{N}=1$ SCFTs
on compact Riemann surfaces with a partial topological twist.
On the field theory side it has been shown that the central charge of the 2d theory can be obtained in terms of the anomalies of the 4d theory through the formalism of the anomaly polynomial.
On the holographic dual side we have shown that this mechanism can be reproduced in terms 
of the central charge of the gauging and of a superpotential.
The flow is triggered in supergravity  by turning on some fluxes for the vectors.
The central charge can be expressed in terms of the fluxes, the isometries and the
sections.
 By using the constraints of the Special geometry we obtained a quadratic formula for the central charge of the 2d theory in terms of the R-charges, representing 
 the mixing coefficients of the vectors with the graviphoton.
  The extremization of the central charge on the field theory side is associated to an attractor mechanism for the scalars on the supergravity side.
 We applied this formalism to a consistent truncation of $T^{1,1}$, containing a Betti vector multiplet, dual to the baryonic symmetry. 
 This allowed us to reproduce the fact that
 the baryonic symmetry does not mix with the R-charge of the 4d case but it can mix
 once the theory is compactified to 2d.
We have shown the existence of solutions interpolating between the AdS$_5$
 and the AdS$_3$ vacua and we have we computed the central charge from the gravitational 
 side, reproducing the field theory results.
We provided further examples of this extremization by matching with the results obtained for other truncations,
corresponding to the partially topologically twisted compactification of  $\mathcal{N}=4$ SYM and of infinite families
of  $\mathcal{N}=1$  SCFTs describing M5 branes on compact Riemann surfaces.

Our analysis can be applied to other truncations in presence of vector multiplets
and hypermultiplets.
For example it may be interesting to study 
the $\mathcal{N}=2$ truncation of 5d maximal gauged supergravity
worked out in \cite{Bobev:2014jva}, corresponding to the holographic dual description
of the $\mathcal{N}=1^*$ Leigh-Strassler fixed point \cite{Leigh:1995ep}.
By reformulating this theory in terms 5d of $\mathcal{N}=2$
gauged supergravity and applying our formalism it should be possible to recover the 2d
central charge obtained by the twisted compactification 
of this model on a hyperbolic Riemann surface.
Another interesting line of research consists in including the effect of higher derivatives,
along the lines of \cite{Baggio:2014hua}. This corresponds to 
including the Pontryagin classes in the calculation of the central charge 
\cite{Benini:2015bwz}, and it is related to sub-leading effects in the rank of the
gauge groups. 

Moreover, the existence of a volume formula for the $\mathcal{M}_7$ manifold
was conjectured in \cite{Benini:2015bwz}.
Such volume formula should reproduce the central charge and its extremization similarly to the 
case of SE$_5$ discussed in \cite{Martelli:2005tp}.
In this paper we obtained an expression for $c_{2d}$ that is quadratic in the sections and 
is formulated in terms of the CS terms of the 5d theory.
The CS terms are associated to triangle anomalies and the volumes of the SE$_5$ geometry \cite{Benvenuti:2006xg}.
This formulation looks a promising starting point to construct the volume formula for the $\mathcal{M}_7$ manifold 
in terms of the geometric data of the SE$_5$ geometry. 

We wish to conclude our discussion by stressing that the connection between the extremization principles 
and the attractor mechanism extends beyond the five-dimensional case.
For example, it has been shown recently in \cite{Benini:2015eyy}  that 
the microstate counting for AdS$_4$ black holes 
can been achieved through the extremization of 
a topologically twisted index on $S^2 \times S^1$.
This extremization principle  
corresponds to the AdS$_2$ attractor. 
It would be interesting to extend the analysis performed here to the case of the
$\mathcal{N}=2$ AdS$_4$ supergravity truncations in presence of Betti multiplets, performed in \cite{Cassani:2012pj}.
Investigation of the extremization principle in such theories 
might shed light on the microscopic origin of the AdS$_4$ BPS black holes found in 
\cite{Halmagyi:2013sla} by exploiting the results found in \cite{Hosseini:2016tor,Hosseini:2016ume}.

\section*{Acknowledgments}
We are grateful to D.~Cassani, L.~Cassia, K.~Hristov, S.~Penati and P.~ Szepietowski for useful comments and discussions, and R.~Monten for help in the analysis of the numerical solutions in Sec. \ref{ads3vacuumsect} and collaboration in related projects.
We acknowledge support from the Simons
Center for Geometry and Physics, Stony Brook University
for hospitality during some steps of this paper. The work of A.~A.~ is supported by the Swiss National Science Foundation 
(snf) under grant number pp00p2$\_$157571/1.
C.~T.~ acknowledges support
from the NWO Rubicon grant, Columbia University and
from DOE grant DE-SC0011941.

\appendix
\section{$\mathcal{N}=2$ 5d gauged supergravity}
\label{AppA}
In this appendix we review some basic aspects of 5d Gauged supergravity and we fix the conventions we use in the main body of the paper for the analysis of the truncation of $T^{1,1}$ with the Betti vector multiplet turned on.  The general matter content of $\mathcal N=2$ supergravity consists in the gravity multiplet, and vector, tensor and hypermultiplets. In the $T^{1,1}$ truncation taken into consideration tensor multiplets do not appear, hence we do not consider them here. We refer the reader to \cite{Ceresole:2000jd} (and \cite{Bergshoeff:2004kh} as well) for a more general discussion of gauged $\mathcal N=2$ 5d supergravity. In what follows we will stick to the conventions used in \cite{Halmagyi:2011yd}.

The bosonic field content of the supergravity theory is as follows. The gravity multiplet consists in the metric $g_{\mu\nu}$ and one vector $A_{\mu}^0$. We have then $n_v$ vector multiplets, which consist in $n_v$ vectors $A_{\mu}^I$, $I=1\ldots,n_v$, and vector multiplet scalars $\phi^x$, living on a Very Special manifold. There are finally $n_h$ hypermultiplets, for a total of $4n_h$ hyperscalars $q^X$ which parameterize a Quaternionic manifold.
The bosonic $\mathcal N=2$ Lagrangian has this form:
\begin{eqnarray}
\mathcal L &=& R - \frac12 g_{xy}D_\mu\phi^x D^\mu\phi^y - \frac12 g_{XY}D_\mu q^X D^\mu q^Y  \nonumber \\
&& - \frac14 G_{IJ}F^I_{\mu\nu}F^{J\,\mu\nu} + \frac1{24}c_{IJK}\epsilon^{\mu\nu\rho\lambda\sigma}F^I_{\mu\nu}F^J_{\rho\lambda}A^K_\sigma - V \,.
\end{eqnarray}
The scalar potential is
\begin{equation}
V = 2 g^{xy} \partial_x X^I \partial_y X^J P_I^r P_J^r - \frac{4}{3} (P_I^r X^I)^2 +\frac{1}{2} g_{XY} K^X_I K^Y_J X^I X^J\,,
\end{equation}
and the covariant derivatives for the vector multiplet 
and for the hypermultiplet scalars (fixing the gauge coupling $g=1$) are defined as
\begin{eqnarray}
D_\mu\phi^x = \partial_\mu\phi^x + A^I_\mu K_I^x(\phi^x)
,\quad \quad
D_\mu q^X = \partial_\mu q^X + A^I_\mu K_I^X(q^X)\,,
\end{eqnarray}
where $K_I^x(\phi^x)$ and $K_I^X(q^X)$ are Killing vectors  corresponding to the gauging of the isometries of the Special manifold and Quaternionic manifold.

The supersymmetry variations
for the gravitino, gauginos and
hyperinos of \cite{Ceresole:2000jd} adapted to our conventions \cite{Halmagyi:2011yd,Liu:2011dw} read%\footnote{Notice that in particular, with respect to \cite{Ceresole:2000jd}, we have $P^r_{theirs}= P^r_{ours}/\sqrt6 $ and $F^{\mu\nu}_{theirs} =\sqrt{3/2} F^{\mu\nu}_{ours}$. See also \cite{Halmagyi:2011yd,Liu:2011dw}.}: 
\begin{eqnarray}
\delta\psi_{\mu\,i}&=&\bigl[D_\mu +\frac{i}{24} X_I(\gamma_\mu{}^{\nu\rho} - 4\delta_\mu^\nu\gamma^\rho)F_{\nu\rho}^I\bigr]\epsilon_i + \frac{i}6 \gamma_{\mu} X^I (P_I)_i{}^j\epsilon_j\,, \nonumber \\
\delta\lambda^x_i &=&\bigl(-\frac{i}2\gamma\cdot D\phi^x - \frac14g^{xy}\partial_y X_I\gamma^{\mu\nu}F_{\mu\nu}^I \bigr)\epsilon_i - g^{xy}\partial_y X^I(P_I)_i{}^j\epsilon_j\,, \nonumber \\
\delta\zeta^A &=& f^{i\,A}_X\bigl(-\frac{i}2\gamma\cdot Dq^X + \frac12 X^I K^X_I\bigr)\epsilon_i\,, 
\label{eq:susy}
\end{eqnarray}
The supercovariant derivative acting on the supersymmetry parameter is
\begin{equation}
D_{\mu} \epsilon_i = \partial_{\mu} \epsilon_i + \frac14 w_{\mu}^{ab} \gamma_{ab} \epsilon_i + \hat{\omega}_{\mu}^x (\sigma^x)_i^j \epsilon_j  +\frac{1}{2} A_{\mu}^I   (P_I)_i{}^j\epsilon_j\,,
\end{equation}
where $\hat{\omega}_{\mu} = \frac{i}{2} \partial_{\mu} q^u \omega_u^x$ is the connection on the Quaternionic manifold.

So in total the susy variation of the gravitino reads:
\begin{eqnarray}
\delta\psi_{\mu\,i} &= &\partial_{\mu} \epsilon_i + \frac14 w_{\mu}^{ab} \gamma_{ab} \epsilon_i  + \hat{\omega}_{\mu}^x (\sigma^x)_i^j \epsilon_j 
 +\frac{1}{2} A_{\mu}^I   (P_I)_i{}^j\epsilon_j +\\
&+&\frac{i}{24} X_I(\gamma_\mu{}^{\nu\rho} - 4\delta_\mu^\nu\gamma^\rho)F_{\nu\rho}^I \epsilon_i + \frac{i}6 X^I \gamma_{\mu} (P_I)_i{}^j\epsilon_j\,.
\end{eqnarray}
The vector multiplet scalars $\phi^x$ are often expressed in terms of the scalars $X^I(\phi^x)$, constrained by $\frac{1}{6} C_{IJK} X^I X^J X^K = 1$.
Moreover, the metric on the scalar manifold is given by 
\begin{equation}
G_{IJ} = X_I X_J - C_{IJK}X^K,
\quad
g_{xy}= \partial_x X^I \partial_y X^J G_{IJ},
\end{equation}
with the $X_I$ fields (lower index) given by
\begin{equation}
X_I = \frac12 c_{IJK} X^J X^K\,.
\end{equation}

The fields $f_i^{A X}$ are the vielbeins on the Quaternionic manifold \cite{Ceresole:2000jd}, with the $SU(2)$ index $i=1,2$ and the $Sp(2n_h)$ index $A=1,...,2n_h$ raised and lowered as usual by the matrices $\varepsilon^{ij}$ and $C_{AB}$. The metric on the Quaternionic manifold is
\be \label{kahmetric}
g_{XY} = f_{X}^{iA} f_{Y}^{jB} \varepsilon_{ij} C_{AB} = f_{X}^{iA} f_{YiA} \,.
\ee
The $SU(2)$ curvature is
\be
\label{su2curvature} \Omega_{XYij} = f_{XC(i}f_{j)Y}^C \,,
\ee
and the connection $\omega_{\mu,i}^j = (\partial_{\mu} q^X) \omega_{X,i}^j$ is such that 
\be
\Omega_{XYi}^j =  i \Omega_{XY}^r (\sigma_r)_i^j
= 2 \partial_{[X} \omega_{Y]i}^j -2 \omega_{[X|i|}^k \omega_{Y]k}^j \,, \ee
\be
\Omega^r = d \omega^r - \varepsilon^{rst} \omega^s \omega^t\,.
\ee
The SU(2) curvatures are proportional to the KyperKahler forms, hence
\be
\Omega^r_{XY}\Omega^{sYZ} = -\frac14 \delta^{rs} \delta_X^Z -\frac12 \varepsilon^{rst} {\Omega_X^t}^Z\,.
\ee
%\begin{eqnarray}
%g_{XY} f_{iA}^X f_{jB}^Y &= & \epsilon_{ij} C_{AB}\\
% f_{iC}^X f_{j}^{YC} + f_{iC}^Y f_{j}^{XC}  &= & g^{XY} \epsilon_{ij} \\
%f_{iA}^X f_{B}^{Yi} +f_{iA}^Y f_{B}^{Xi} &= & \frac{1}{n_H} g^{XY}  C_{AB}
%\end{eqnarray}
Finally, the Killing prepotentials $(P_I)_i^j = P_I^r(i \sigma^r)_i^j$ are related to the Killing vectors $K_I^X$ by the equation
\begin{equation}
K_I^X \Omega_{XY}^r = D_Y P_I^r\,, \qquad D_X P^r_I \equiv\partial_X P_I^r + 2 \varepsilon^{rst} \omega^s_X P^t_I\,.
\end{equation}
It is useful to define the following quantities:
\be
P^r \equiv X^I P_I^r\,, \qquad K^X \equiv X^I K_I^X\,.
\ee
We moreover decompose the vector $P^r$ in its norm and phase 
\be \label{phase}
P^r =  W Q^r\,, \qquad Q^r Q^r =1\,, \qquad W = \sqrt{ P^r P^r}
\,.
\ee
Lastly, one can straightforwardly verify the relation
\be\label{derpot}
\partial_{X} W = \frac1W P^r D_X P_r = \frac1W P^r \Omega^r_{XY} K^Y
\ee
which will be useful later.

\section{BPS equations }
\label{variation}
We give here a detailed derivation of the BPS equations, which are obtained by setting to zero the supersymmetry variations of the gravitino, gaugino and hyperino. In doing so, we plug the ansatz of the metric \eqref{metric}\eqref{acca} and purely magnetic gauge field $A_{x}$ and $A_y$ (while $A_t=0$) in the supersymmetry variations \eqref{eq:susy}. We make the following ansatz for the Killing spinor projection relations, where the numbers indicate flat indices:
\be \label{proj}
\gamma_2  \epsilon_i = - Q^r  {\sigma^r}_{ij} \epsilon^j \,, \qquad \gamma_{34} \epsilon_i =- i Q^r  {\sigma^r}_{ij} \epsilon^j  \,, \qquad \partial_t \epsilon=\partial_z \epsilon=\partial_{\text{x}} \epsilon=\partial_{\text{y}} \epsilon =0\,.
\ee
Imposing these projectors amount to preserving $1/4$ of the supersymmetries throughout the flow. For the solutions taken into consideration, there is a supersymmetry enhancement at the IR: the $AdS_3 \times \Sigma_g$ geometry is in fact 1/2 BPS - see discussion in sec. \ref{ads3vacuumsect}. 

We will make the following choice of vielbeins \be
e_t^0 = e_z^1 =e_r^2 = e^{f(r)} \,, \qquad e_{\text{x}}^3=e^{g(r)} \qquad  e_{\text{y}}^4 = e^{g(r)}F(\text{x})\,, 
\ee
where the function $F$ is defined in \eqref{acca}.
The non vanishing components of the spin connection are:
\be
w_t^{02} =w_z^{12}= f' \,, \qquad w_{\text{x}}^{23} =-g' e^{g-f}\,, \qquad   w_{\text{y}}^{24} = -g' e^{g-f} F(\text{x})\,, \qquad w_{\text{y}}^{34}  = -F'(\text{x}) \,.
\ee
The analysis which follows apply to the case $k = \pm1$, but the procedure can be straightforwardly applied to the $\kappa=0$ case as well.
\subsection{Gravitino}
 
The BPS equation derived from the $t$ component of the gravitino variation $\delta \psi_{ti}$ reads
\begin{equation}
\partial_t\epsilon_i +\frac14 w_t^{02} \gamma_{02} \epsilon_i +\frac{i}{24} X_I \gamma_{t}^{\text{xy}} F_{\text{xy}}^I \epsilon_i +\frac{i}{6} \gamma_{t} X^I {P_I}_i^j \epsilon_j =0\,.
\end{equation}
We impose now the relations \eqref{proj}. Therefore, the first term drops out, and the other give the equation
\be
-\frac12 f' \gamma_0 Q^r  {\sigma^r}_{ij} \epsilon^j  -\frac{X_I a^I}{12} e^{-2g +f}\gamma_0   Q^r  {\sigma^r}_{ij} \epsilon^j  -\frac{1}{6} \gamma_0 e^f P^r {\sigma^r}_{ij} \epsilon^j =0\,,
\ee
which, given \eqref{phase}, can be simplified in
\be
f'+ \frac{1}{6}a^I X_I e^{-2g+f}+\frac{1}3 e^{f} W =0 \label{eqt}\,.
\ee

From the $r$-component, using \eqref{proj} we have
\begin{equation}
\partial_r \epsilon_i + \hat{\omega}_{\mu}^x (\sigma^x)_i^j \epsilon_j- \frac{i}{24} a^I X_I e^{-2g +f} \gamma_2^{34} \epsilon_i +\frac{i}{6} \gamma_{2} e^f X^I {P_I}_i^j \epsilon_j =0\,.
\end{equation}
which determines the functional form of in terms of $r$ of the Killing spinor $\epsilon_i(r)$. 
%\be
%\epsilon_i(r)= e^{\int  ( \partial_r q^X \omega_{X1}^j  \epsilon_j- \frac{1}{12} a^I X_I e^{-2g +f} -\frac{i}{6} e^f W) dr} \epsilon_{0,i}\,.
%\ee
%We will give the explicit form of the radial dependence of when solving for the explicit model. \eqref{proj}.

The $\text{x}$ component of the gravitino variation gives:
\begin{equation}
\partial_{\text{x}} \epsilon_i +\frac14 w_{\text{x}}^{23} \gamma_{23} \epsilon_i  -\frac{i}{6} \gamma_{\text{y}}   a^I  X_I \epsilon_i+\frac{i}{6} \gamma_{\text{x}} X^I {P_I}_i^j \epsilon_j =0 \nonumber\,,
\end{equation}
which after using \eqref{proj} yields
\be \label{xalsoy}
g' -\frac{1}{3} X_I a^{I } e^{-2g+f}  + \frac{1}3 e^f W   =0 \,.
\ee

The $\text{y}$ component of the gravitino variation gives:
\begin{equation}
\partial_{\text{y}} \epsilon_i +\frac14 w_{\text{y}}^{23} \gamma_{23} \epsilon_i +\frac14 w_{\text{y}}^{34} \gamma_{34} \epsilon_i + \frac{1}{2} A_{\text{y}}^I (P_I)_i^j \epsilon_j  -\frac{i}{6} \gamma_{\text{x}}   a^I  X_I \epsilon_i+\frac{i}{6} \gamma_{\text{y}} X^I {P_I}_i^j \epsilon_j =0 \,.\nonumber
\end{equation}
Imposing the projection relations \eqref{proj}, we obtain two parts, one multiplying  $Q^r  {\sigma^r}_{ij} $ and the other $ \gamma_{\text{y}} Q^r  {\sigma^r}_{ij} $, that need to vanish separately. From the terms in $ \gamma_{\text{y}} Q^r  {\sigma^r}_{ij} $ we get exactly  equation \eqref{xalsoy} obtained before.
The part multiplying ${P_I}_i^j \epsilon_j$ yields
\be
 \frac{1}{2} A_{\text{y}}^I P_I^r =  \frac{\kappa}{2}Q^r F'(\text{x}) \qquad \rightarrow \qquad a^I P_I^r  = -\kappa  Q^r\,,
\ee
where we made use of the relations \eqref{gaugfield}.

 Finally, the $z$ component of the gravitino variation gives exactly the same equation \eqref{eqt}.

\subsection{Gaugino}
After plugging in the ansatz \eqref{metric}\eqref{acca}\eqref{AnsatzRsugra}, and retaining only the radial dependence of the fields, the gaugino equation reads
\be
-\frac{i}{2} \gamma^r  \, \partial_r \phi^x \, \epsilon_i -\frac{i}{2} \gamma^{\phi} \, a^I K_{I}^x\, \epsilon_i -\frac{i}{2} \gamma^{\theta} \, a^I K_{I}^x\, \epsilon_i
+\frac12 g^{xy} \partial_y X_I \gamma^{\text{x} \text{y}} a^I \epsilon_i -\frac12g^{xy} \partial_{y} X^I (P_I)_i^j \epsilon_j =0\,.
\ee
Since the models taken into consideration don't gauge the isometries of the Special manifold, we have $K_I^x=0$. Hence, after using the projections \eqref{proj} we are left with the equations 
\be
\frac12 \partial_r \phi^x Q^r\,  +
\frac12 e^{-2g+f} g^{xy} \partial_y X_I  a^I Q^r - e^f\, g^{xy} \partial_{y} X^I P_I^r =0\,.
\ee

\subsection{Hyperino}
The hyperino equation instead requires a bit more effort. Plugging in all components, and taking into account the radial dependence of the scalar fields, it reads
\be
f_X^{iA} \left( -\frac{i}{2}\gamma^r \partial_r q^X + \gamma^{\text{y}}A_{\text{y}}^I K_I^X  +\frac12 X^I K_I^X \right) \epsilon_i=0\,.
\ee
The second and third term should vanish separately, hence the constraint 
\be
 a^I K_I^X =0\,.
\ee
We are now ready to massage the remaining equation
\be \label{eqhyper}
f_X^{iA} \left( -i \gamma^r \partial_r q^X  +X^I K_I^X \right) \epsilon_i=0\,,
\ee
following the procedure spelled out in \cite{Ceresole:2001wi}.  We multiply the left hand side of this equations by $f_{Aj}^Y$. Using the relations \eqref{kahmetric} and \eqref{su2curvature} we get
\be
f_{XjA} f_Y^{iA} = \frac12 g_{XY} \delta_i^j +{\Omega_{XYj}}^i
\ee
so that eq. \eqref{eqhyper}, using the projection relations \eqref{proj}, becomes of the form
\be
\left[ g_{XY} \delta_j^i +2i {\Omega_{XY}}^r {{\sigma^r}_j}^i
\right] [i q^{X'} Q^s {{\sigma^s}_i}^k +  K^X \delta_i^k] =0\,,
\ee
also written as a matrix equation
\be
[A_Y \delta_j^k+ i B_Y^r (\sigma^r)_j^k] \epsilon_k =0\,,
\ee
where 
\be
A_Y=  K_Y - 2\Omega_{YX}^r Q^r q^{X'}\,,
\ee
\be
B_Y^r= g_{YX} q^{X'}  Q^r +  \varepsilon^{rst} \, \Omega_{YX}^s Q^t q^{X'} +  2 \, \Omega_{YX}^r K^X\,.
\ee
Both $A_Y$ and $B_Y^r$ need to vanish, and they do if the following condition is enforced  (more details in \cite{Ceresole:2001wi})

\be
q^{Y'} g_{YX} =  2 Q^r{ \Omega^r}_{XY} K^Y
\ee
where $W$ is defined in \eqref{phase} and we have used \eqref{derpot}. This is the last BPS equation.
\vspace{5mm}

\noindent To sum up, we have massaged the BPS equations in the neat system of boson equations
\begin{eqnarray}
f'+ \frac{1}3 e^{f} W + \frac{1}{6}a^I X_I e^{-2g+f}&=&0 \\
g' + \frac{1}3 e^f W-\frac{1}{3} a^I X_I e^{-2g+f} &=&0 \\ 
\frac12\partial_r \phi^x Q^r\,  +
\frac12 e^{-2g+f} g^{xy} \partial_y X_I  a^I Q^r - e^f\, g^{xy} \partial_{y} X^I P_I^r & = & 0 \\
q^{Y'} g_{YX} -  2 Q^r{ \Omega^r}_{XY} K^Y & = & 0 \\
a^I P_I^r  +\kappa  Q^r & = & 0\\
 a^I K_I^X& =& 0\,.
\end{eqnarray}
Given these equations, the Maxwell's and Bianchi equations are automatically satisfied for the choice of warp factors and gauge fields  done in section \ref{sec:conifoldBetti}.

\bibliographystyle{JHEP}
\bibliography{references}

\providecommand{\href}[2]{#2}\begingroup\raggedright\begin{thebibliography}{10}

\bibitem{Witten:1988xj}
E.~Witten, {\it {Topological Sigma Models}},  {\em Commun. Math. Phys.} {\bf
  118} (1988) 411.

\bibitem{Bershadsky:1995vm}
M.~Bershadsky, A.~Johansen, V.~Sadov, and C.~Vafa, {\it {Topological reduction
  of 4-d SYM to 2-d sigma models}},  {\em Nucl. Phys.} {\bf B448} (1995)
  166--186, [\href{http://arxiv.org/abs/hep-th/9501096}{{\tt hep-th/9501096}}].

\bibitem{Bershadsky:1995qy}
M.~Bershadsky, C.~Vafa, and V.~Sadov, {\it {D-branes and topological field
  theories}},  {\em Nucl. Phys.} {\bf B463} (1996) 420--434,
  [\href{http://arxiv.org/abs/hep-th/9511222}{{\tt hep-th/9511222}}].

\bibitem{Maldacena:2000mw}
J.~M. Maldacena and C.~Nunez, {\it {Supergravity description of field theories
  on curved manifolds and a no go theorem}},  {\em Int. J. Mod. Phys.} {\bf
  A16} (2001) 822--855, [\href{http://arxiv.org/abs/hep-th/0007018}{{\tt
  hep-th/0007018}}]. [,182(2000)].

\bibitem{Johansen:2003hw}
A.~Johansen, {\it {Holomorphic currents and duality in N = 1 supersymmetric
  theories}},  {\em JHEP} {\bf 12} (2003) 032,
  [\href{http://arxiv.org/abs/hep-th/0309125}{{\tt hep-th/0309125}}].

\bibitem{Kapustin:2006hi}
A.~Kapustin, {\it {Holomorphic reduction of N=2 gauge theories, Wilson-'t Hooft
  operators, and S-duality}},  \href{http://arxiv.org/abs/hep-th/0612119}{{\tt
  hep-th/0612119}}.

\bibitem{Donos:2008ug}
A.~Donos, J.~P. Gauntlett, and N.~Kim, {\it {AdS Solutions Through
  Transgression}},  {\em JHEP} {\bf 09} (2008) 021,
  [\href{http://arxiv.org/abs/0807.4375}{{\tt arXiv:0807.4375}}].

\bibitem{Benini:2013cda}
F.~Benini and N.~Bobev, {\it {Two-dimensional SCFTs from wrapped branes and
  c-extremization}},  {\em JHEP} {\bf 06} (2013) 005,
  [\href{http://arxiv.org/abs/1302.4451}{{\tt arXiv:1302.4451}}].

\bibitem{Donos:2014eua}
A.~Donos and J.~P. Gauntlett, {\it {Flowing from AdS$_{5}$ to AdS$_{3}$ with
  T$^{1,1}$}},  {\em JHEP} {\bf 08} (2014) 006,
  [\href{http://arxiv.org/abs/1404.7133}{{\tt arXiv:1404.7133}}].

\bibitem{Baggio:2014hua}
M.~Baggio, N.~Halmagyi, D.~R. Mayerson, D.~Robbins, and B.~Wecht, {\it {Higher
  Derivative Corrections and Central Charges from Wrapped M5-branes}},  {\em
  JHEP} {\bf 12} (2014) 042, [\href{http://arxiv.org/abs/1408.2538}{{\tt
  arXiv:1408.2538}}].

\bibitem{Bea:2015fja}
Y.~Bea, J.~D. Edelstein, G.~Itsios, K.~S. Kooner, C.~Nunez, D.~Schofield, and
  J.~A. Sierra-Garcia, {\it {Compactifications of the Klebanov-Witten CFT and
  new AdS$_{3}$ backgrounds}},  {\em JHEP} {\bf 05} (2015) 062,
  [\href{http://arxiv.org/abs/1503.07527}{{\tt arXiv:1503.07527}}].

\bibitem{Benini:2015bwz}
F.~Benini, N.~Bobev, and P.~M. Crichigno, {\it {Two-dimensional SCFTs from
  D3-branes}},  {\em JHEP} {\bf 07} (2016) 020,
  [\href{http://arxiv.org/abs/1511.09462}{{\tt arXiv:1511.09462}}].

\bibitem{Schafer-Nameki:2016cfr}
S.~Schafer-Nameki and T.~Weigand, {\it {F-theory and 2d $(0, 2)$ theories}},
  {\em JHEP} {\bf 05} (2016) 059, [\href{http://arxiv.org/abs/1601.02015}{{\tt
  arXiv:1601.02015}}].

\bibitem{Apruzzi:2016iac}
F.~Apruzzi, F.~Hassler, J.~J. Heckman, and I.~V. Melnikov, {\it {UV Completions
  for Non-Critical Strings}},  {\em JHEP} {\bf 07} (2016) 045,
  [\href{http://arxiv.org/abs/1602.04221}{{\tt arXiv:1602.04221}}].

\bibitem{Apruzzi:2016nfr}
F.~Apruzzi, F.~Hassler, J.~J. Heckman, and I.~V. Melnikov, {\it {From 6D SCFTs
  to Dynamic GLSMs}},  \href{http://arxiv.org/abs/1610.00718}{{\tt
  arXiv:1610.00718}}.

\bibitem{Kim:2005ez}
N.~Kim, {\it {AdS(3) solutions of IIB supergravity from D3-branes}},  {\em
  JHEP} {\bf 01} (2006) 094, [\href{http://arxiv.org/abs/hep-th/0511029}{{\tt
  hep-th/0511029}}].

\bibitem{Figueras:2007cn}
P.~Figueras, O.~A.~P. Mac~Conamhna, and E.~O~Colgain, {\it {Global geometry of
  the supersymmetric AdS(3)/CFT(2) correspondence in M-theory}},  {\em Phys.
  Rev.} {\bf D76} (2007) 046007,
  [\href{http://arxiv.org/abs/hep-th/0703275}{{\tt hep-th/0703275}}].

\bibitem{Benvenuti:2004dy}
S.~Benvenuti, S.~Franco, A.~Hanany, D.~Martelli, and J.~Sparks, {\it {An
  Infinite family of superconformal quiver gauge theories with Sasaki-Einstein
  duals}},  {\em JHEP} {\bf 06} (2005) 064,
  [\href{http://arxiv.org/abs/hep-th/0411264}{{\tt hep-th/0411264}}].

\bibitem{Gauntlett:2004yd}
J.~P. Gauntlett, D.~Martelli, J.~Sparks, and D.~Waldram, {\it {Sasaki-Einstein
  metrics on S**2 x S**3}},  {\em Adv. Theor. Math. Phys.} {\bf 8} (2004),
  no.~4 711--734, [\href{http://arxiv.org/abs/hep-th/0403002}{{\tt
  hep-th/0403002}}].

\bibitem{Intriligator:2003jj}
K.~A. Intriligator and B.~Wecht, {\it {The Exact superconformal R symmetry
  maximizes a}},  {\em Nucl. Phys.} {\bf B667} (2003) 183--200,
  [\href{http://arxiv.org/abs/hep-th/0304128}{{\tt hep-th/0304128}}].

\bibitem{Bertolini:2004xf}
M.~Bertolini, F.~Bigazzi, and A.~L. Cotrone, {\it {New checks and subtleties
  for AdS/CFT and a-maximization}},  {\em JHEP} {\bf 12} (2004) 024,
  [\href{http://arxiv.org/abs/hep-th/0411249}{{\tt hep-th/0411249}}].

\bibitem{Butti:2005vn}
A.~Butti and A.~Zaffaroni, {\it {R-charges from toric diagrams and the
  equivalence of a-maximization and Z-minimization}},  {\em JHEP} {\bf 11}
  (2005) 019, [\href{http://arxiv.org/abs/hep-th/0506232}{{\tt
  hep-th/0506232}}].

\bibitem{Schmude:2016bqp}
J.~Schmude and O.~Vasilakis, {\it {Superconformal Symmetry in the Kaluza-Klein
  Spectrum of Warped AdS(3)}},  \href{http://arxiv.org/abs/1605.00636}{{\tt
  arXiv:1605.00636}}.

\bibitem{Klebanov:1998hh}
I.~R. Klebanov and E.~Witten, {\it {Superconformal field theory on three-branes
  at a Calabi-Yau singularity}},  {\em Nucl. Phys.} {\bf B536} (1998) 199--218,
  [\href{http://arxiv.org/abs/hep-th/9807080}{{\tt hep-th/9807080}}].

\bibitem{Cassani:2010na}
D.~Cassani and A.~F. Faedo, {\it {A Supersymmetric consistent truncation for
  conifold solutions}},  {\em Nucl. Phys.} {\bf B843} (2011) 455--484,
  [\href{http://arxiv.org/abs/1008.0883}{{\tt arXiv:1008.0883}}].

\bibitem{Bena:2010pr}
I.~Bena, G.~Giecold, M.~Grana, N.~Halmagyi, and F.~Orsi, {\it {Supersymmetric
  Consistent Truncations of IIB on $T^{1,1}$}},  {\em JHEP} {\bf 04} (2011)
  021, [\href{http://arxiv.org/abs/1008.0983}{{\tt arXiv:1008.0983}}].

\bibitem{Karndumri:2013iqa}
P.~Karndumri and E.~O~Colg\'ain, {\it {Supergravity dual of c-extremization}},
  {\em Phys. Rev.} {\bf D87} (2013), no.~10 101902,
  [\href{http://arxiv.org/abs/1302.6532}{{\tt arXiv:1302.6532}}].

\bibitem{Bah:2012dg}
I.~Bah, C.~Beem, N.~Bobev, and B.~Wecht, {\it {Four-Dimensional SCFTs from
  M5-Branes}},  {\em JHEP} {\bf 06} (2012) 005,
  [\href{http://arxiv.org/abs/1203.0303}{{\tt arXiv:1203.0303}}].

\bibitem{Klemm:2016kxw}
D.~Klemm, N.~Petri, and M.~Rabbiosi, {\it {Black string first order flow in
  N=2, d=5 abelian gauged supergravity}},
  \href{http://arxiv.org/abs/1610.07367}{{\tt arXiv:1610.07367}}.

\bibitem{Tachikawa:2005tq}
Y.~Tachikawa, {\it {Five-dimensional supergravity dual of a-maximization}},
  {\em Nucl. Phys.} {\bf B733} (2006) 188--203,
  [\href{http://arxiv.org/abs/hep-th/0507057}{{\tt hep-th/0507057}}].

\bibitem{Barnes:2005bw}
E.~Barnes, E.~Gorbatov, K.~A. Intriligator, and J.~Wright, {\it {Current
  correlators and AdS/CFT geometry}},  {\em Nucl. Phys.} {\bf B732} (2006)
  89--117, [\href{http://arxiv.org/abs/hep-th/0507146}{{\tt hep-th/0507146}}].

\bibitem{Brown:1986nw}
J.~D. Brown and M.~Henneaux, {\it {Central Charges in the Canonical Realization
  of Asymptotic Symmetries: An Example from Three-Dimensional Gravity}},  {\em
  Commun. Math. Phys.} {\bf 104} (1986) 207--226.

\bibitem{deWit:2003ja}
B.~de~Wit, I.~Herger, and H.~Samtleben, {\it {Gauged locally supersymmetric D =
  3 nonlinear sigma models}},  {\em Nucl. Phys.} {\bf B671} (2003) 175--216,
  [\href{http://arxiv.org/abs/hep-th/0307006}{{\tt hep-th/0307006}}].

\bibitem{Karndumri:2013dca}
P.~Karndumri and E.~O. Colg\'ain, {\it {3D Supergravity from wrapped
  D3-branes}},  {\em JHEP} {\bf 10} (2013) 094,
  [\href{http://arxiv.org/abs/1307.2086}{{\tt arXiv:1307.2086}}].

\bibitem{Amariti:2015ybz}
A.~Amariti and A.~Gnecchi, {\it {3D $\tau_{RR}$-minimization in AdS$_4$ gauged
  supergravity}},  {\em JHEP} {\bf 07} (2016) 006,
  [\href{http://arxiv.org/abs/1511.08214}{{\tt arXiv:1511.08214}}].

\bibitem{Benini:2015eyy}
F.~Benini, K.~Hristov, and A.~Zaffaroni, {\it {Black hole microstates in
  AdS$_{4}$ from supersymmetric localization}},  {\em JHEP} {\bf 05} (2016)
  054, [\href{http://arxiv.org/abs/1511.04085}{{\tt arXiv:1511.04085}}].

\bibitem{Benini:2015noa}
F.~Benini and A.~Zaffaroni, {\it {A topologically twisted index for
  three-dimensional supersymmetric theories}},  {\em JHEP} {\bf 07} (2015) 127,
  [\href{http://arxiv.org/abs/1504.03698}{{\tt arXiv:1504.03698}}].

\bibitem{Buchel:2006gb}
A.~Buchel and J.~T. Liu, {\it {Gauged supergravity from type IIB string theory
  on Y**p,q manifolds}},  {\em Nucl. Phys.} {\bf B771} (2007) 93--112,
  [\href{http://arxiv.org/abs/hep-th/0608002}{{\tt hep-th/0608002}}].

\bibitem{Louis:2012ux}
J.~Louis, P.~Smyth, and H.~Triendl, {\it {Supersymmetric Vacua in N=2
  Supergravity}},  {\em JHEP} {\bf 08} (2012) 039,
  [\href{http://arxiv.org/abs/1204.3893}{{\tt arXiv:1204.3893}}].

\bibitem{Ceresole:1999ht}
A.~Ceresole, G.~Dall'Agata, and R.~D'Auria, {\it {K K spectroscopy of type IIB
  supergravity on AdS(5) x T**11}},  {\em JHEP} {\bf 11} (1999) 009,
  [\href{http://arxiv.org/abs/hep-th/9907216}{{\tt hep-th/9907216}}].

\bibitem{Halmagyi:2011yd}
N.~Halmagyi, J.~T. Liu, and P.~Szepietowski, {\it {On N = 2 Truncations of IIB
  on T$^{1,1}$}},  {\em JHEP} {\bf 07} (2012) 098,
  [\href{http://arxiv.org/abs/1111.6567}{{\tt arXiv:1111.6567}}].

\bibitem{Halmagyi:2013sla}
N.~Halmagyi, M.~Petrini, and A.~Zaffaroni, {\it {BPS black holes in $AdS_{4}$
  from M-theory}},  {\em JHEP} {\bf 08} (2013) 124,
  [\href{http://arxiv.org/abs/1305.0730}{{\tt arXiv:1305.0730}}].

\bibitem{Herzog:2009gd}
C.~P. Herzog, I.~R. Klebanov, S.~S. Pufu, and T.~Tesileanu, {\it {Emergent
  Quantum Near-Criticality from Baryonic Black Branes}},  {\em JHEP} {\bf 03}
  (2010) 093, [\href{http://arxiv.org/abs/0911.0400}{{\tt arXiv:0911.0400}}].

\bibitem{Bobev:2014jva}
N.~Bobev, K.~Pilch, and O.~Vasilakis, {\it {(0, 2) SCFTs from the
  Leigh-Strassler fixed point}},  {\em JHEP} {\bf 06} (2014) 094,
  [\href{http://arxiv.org/abs/1403.7131}{{\tt arXiv:1403.7131}}].

\bibitem{Karndumri:2015sia}
P.~Karndumri and E.~O. Colgain, {\it {3D supergravity from wrapped M5-branes}},
   {\em JHEP} {\bf 03} (2016) 188, [\href{http://arxiv.org/abs/1508.00963}{{\tt
  arXiv:1508.00963}}].

\bibitem{Szepietowski:2012tb}
P.~Szepietowski, {\it {Comments on a-maximization from gauged supergravity}},
  {\em JHEP} {\bf 12} (2012) 018, [\href{http://arxiv.org/abs/1209.3025}{{\tt
  arXiv:1209.3025}}].

\bibitem{Gaiotto:2009we}
D.~Gaiotto, {\it {N=2 dualities}},  {\em JHEP} {\bf 08} (2012) 034,
  [\href{http://arxiv.org/abs/0904.2715}{{\tt arXiv:0904.2715}}].

\bibitem{Henningson:1998gx}
M.~Henningson and K.~Skenderis, {\it {The Holographic Weyl anomaly}},  {\em
  JHEP} {\bf 07} (1998) 023, [\href{http://arxiv.org/abs/hep-th/9806087}{{\tt
  hep-th/9806087}}].

\bibitem{Leigh:1995ep}
R.~G. Leigh and M.~J. Strassler, {\it {Exactly marginal operators and duality
  in four-dimensional N=1 supersymmetric gauge theory}},  {\em Nucl. Phys.}
  {\bf B447} (1995) 95--136, [\href{http://arxiv.org/abs/hep-th/9503121}{{\tt
  hep-th/9503121}}].

\bibitem{Martelli:2005tp}
D.~Martelli, J.~Sparks, and S.-T. Yau, {\it {The Geometric dual of
  a-maximisation for Toric Sasaki-Einstein manifolds}},  {\em Commun. Math.
  Phys.} {\bf 268} (2006) 39--65,
  [\href{http://arxiv.org/abs/hep-th/0503183}{{\tt hep-th/0503183}}].

\bibitem{Benvenuti:2006xg}
S.~Benvenuti, L.~A. Pando~Zayas, and Y.~Tachikawa, {\it {Triangle anomalies
  from Einstein manifolds}},  {\em Adv. Theor. Math. Phys.} {\bf 10} (2006),
  no.~3 395--432, [\href{http://arxiv.org/abs/hep-th/0601054}{{\tt
  hep-th/0601054}}].

\bibitem{Cassani:2012pj}
D.~Cassani, P.~Koerber, and O.~Varela, {\it {All homogeneous N=2 M-theory
  truncations with supersymmetric AdS4 vacua}},  {\em JHEP} {\bf 11} (2012)
  173, [\href{http://arxiv.org/abs/1208.1262}{{\tt arXiv:1208.1262}}].

\bibitem{Hosseini:2016tor}
S.~M. Hosseini and A.~Zaffaroni, {\it {Large $N$ matrix models for 3d ${\cal
  N}=2$ theories: twisted index, free energy and black holes}},  {\em JHEP}
  {\bf 08} (2016) 064, [\href{http://arxiv.org/abs/1604.03122}{{\tt
  arXiv:1604.03122}}].

\bibitem{Hosseini:2016ume}
S.~M. Hosseini and N.~Mekareeya, {\it {Large $N$ topologically twisted index:
  necklace quivers, dualities, and Sasaki-Einstein spaces}},  {\em JHEP} {\bf
  08} (2016) 089, [\href{http://arxiv.org/abs/1604.03397}{{\tt
  arXiv:1604.03397}}].

\bibitem{Ceresole:2000jd}
A.~Ceresole and G.~Dall'Agata, {\it {General matter coupled N=2, D = 5 gauged
  supergravity}},  {\em Nucl. Phys.} {\bf B585} (2000) 143--170,
  [\href{http://arxiv.org/abs/hep-th/0004111}{{\tt hep-th/0004111}}].

\bibitem{Bergshoeff:2004kh}
E.~Bergshoeff, S.~Cucu, T.~de~Wit, J.~Gheerardyn, S.~Vandoren, and
  A.~Van~Proeyen, {\it {N = 2 supergravity in five-dimensions revisited}},
  {\em Class. Quant. Grav.} {\bf 21} (2004) 3015--3042,
  [\href{http://arxiv.org/abs/hep-th/0403045}{{\tt hep-th/0403045}}]. [Class.
  Quant. Grav.23,7149(2006)].

\bibitem{Liu:2011dw}
J.~T. Liu and P.~Szepietowski, {\it {Supersymmetry of consistent massive
  truncations of IIB supergravity}},  {\em Phys. Rev.} {\bf D85} (2012) 126010,
  [\href{http://arxiv.org/abs/1103.0029}{{\tt arXiv:1103.0029}}].

\bibitem{Ceresole:2001wi}
A.~Ceresole, G.~Dall'Agata, R.~Kallosh, and A.~Van~Proeyen, {\it
  {Hypermultiplets, domain walls and supersymmetric attractors}},  {\em Phys.
  Rev.} {\bf D64} (2001) 104006,
  [\href{http://arxiv.org/abs/hep-th/0104056}{{\tt hep-th/0104056}}].

\end{thebibliography}\endgroup

\end{document}